\documentclass{aastex}
\usepackage{emulateapj5}

\def\sax{{\it BeppoSAX\,}}
\def\asca{{\it ASCA\,}}
\def\rosat{{\it ROSAT\,}}
\def\rxte{{\it RXTE\,}}
\def\uly{{\it Ulysses\,}}
\def\wind{{\it Wind\,}}

\newcommand{\pfg}{\vspace{0.3cm}}
\newcommand{\PSbox}[3]{\mbox{\includegraphics{#1}\hspace{#2}\rule{0pt}{#3}}}

\shorttitle{RXTE/ASM GRB Light Curves}
\shortauthors{Smith et al.}

\slugcomment{Draft: \today}

\begin{document}
 
\title{X-Ray Light Curves of Gamma-ray Bursts Detected with the All-Sky
Monitor on \rxte}

\author{D. A. Smith\altaffilmark{1}, A. Levine\altaffilmark{2}, H. Bradt\altaffilmark{2,3}, K. Hurley\altaffilmark{4}, M. Feroci\altaffilmark{5}, P. Butterworth\altaffilmark{6,7}, S. Golenetskii\altaffilmark{8}, G. Pendleton\altaffilmark{9,10}, \& S. Phengchamnan\altaffilmark{10}}
\altaffiltext{1}{The University of Michigan, Ann Arbor, MI, 48109}
\altaffiltext{2}{Center for Space Research, Massachusetts Institute of Technology, Cambridge, MA, 02139}
\altaffiltext{3}{Department of Physics, Massachusetts Institute of Technology, Cambridge, MA, 02139}
\altaffiltext{4}{Space Sciences Laboratory, University of California, Berkeley, CA, 94720}
\altaffiltext{5}{Istituto de Astrofisica Spaziale (CNR), 00133, Rome Italy}
\altaffiltext{6}{NASA/Goddard Space Flight Center, Greenbelt, MD, 20771}
\altaffiltext{7}{Emergent Information Technologies, Largo, MD, 20774}
\altaffiltext{8}{Ioffe Physico-Technical Institute, Russian Academy of Sciences, St. Petersburg, 194021, Russia}
\altaffiltext{9}{COLSA Corporation, Huntsville, AL, 35806}
\altaffiltext{10}{Department of Physics, University of Alabama, Huntsville, AL, 35899}

\email{donaldas@umich.edu}

\begin{abstract}

We present X-ray light curves (1.5--12~keV) for fifteen gamma-ray
bursts (GRBs) detected by the All-Sky Monitor on the {\it Rossi X-ray
Timing Explorer}.  We compare these soft X-ray light curves with count
rate histories obtained by the high-energy ($>12$~keV) experiments
BATSE, Konus-\wind, the \sax~Gamma-Ray Burst Monitor, and the burst
monitor on \uly.  We discuss these light curves within the context of
a simple relativistic fireball and synchrotron shock paradigm, and we
address the possibility of having observed the transition between a
GRB and its afterglow.  The light curves show diverse morphologies,
with striking differences between energy bands.  In several bursts,
intervals of significant emission are evident in the ASM energy range
with little or no corresponding emission apparent in the high-energy
light curves.  For example, the final peak of GRB~970815 as recorded
by the ASM is only detected in the softest BATSE energy bands.  We
also study the duration of bursts as a function of energy.  Simple,
singly-peaked bursts seem consistent with the $E^{-0.5}$ power law
expected from an origin in synchrotron radiation, but durations of
bursts that exhibit complex temporal structure are not consistent with
this prediction.  Bursts such as GRB~970828 that show many short
spikes of emission at high energies last significantly longer at low
energies than the synchrotron cooling law would predict.

\end{abstract}


\section{INTRODUCTION\label{sec:intro}}

\subsection{X-Rays from Gamma-Ray Bursts}

The first detection of a gamma-ray burst (GRB) at X-ray energies
($\sim$1--15~keV) was made in 1972 with two proportional counters on
the {\it OSO-7} satellite~\citep{wubde73}.  For the next 15 years,
there were few X-ray observations of GRBs, but they showed that the
investigation of a new region of the spectrum could reveal interesting
GRB properties.  Instruments on {\it Apollo-16} were used to detect a
GRB, also in 1972, and light curves in several energy bands showed
that the spectrum evolved over the burst's single
peak~\citep{mpgpt74,tesam74}.  Four bursts were detected with the Air
Force satellite {\it P78-1} in 1979, and the peak X-ray emission was
found to lag the peak gamma-ray emission~\citep{lefks84}.

Analysis of $\sim150$ bursts detected by the Konus instruments on {\it
Venera 11} and {\it Venera 12} showed spectral evolution to be the
rule rather than the exception~\citep{maz81}.  \citet{gmai83} found
that the spectral evolution of a small sample of GRBs could be
characterized in terms of a hardness--intensity correlation.
Later,~\citet{nsmdd86} reported a tendency for high-energy emission to
lead the low-energy emission in bursts detected by instruments on the
{\it Solar Maximum Mission} satellite. \citet{fbmbp95} extended the
study of spectral evolution to BATSE observations of GRBs and found
that the peak energy of the spectral distribution increases
concurrently with or slightly ahead of major increases in source
intensity.  They also found that when multiple peaks are present,
later peaks tend to be softer than earlier peaks.  A more extensive
discussion of spectral evolution in GRBs may be found in~\citet{lp00}.

Studies of X-ray counterparts to GRBs intensified in the late 1980s.
The {\it Ginga} satellite was equipped with co-aligned wide-field
detectors to cover the energy range from 2--400~keV~\citep{mfhin89}.
These instruments were used to detect $\sim$120 bursts between 1987
and 1991~\citep{omnyf91}.  Analysis of twenty-two of these
observations confirmed that spectral softening is common in the tails
of bursts and showed that the X-ray band can contain a large fraction
of the energy emitted from a GRB~\citep{sfmy98}.  X-ray precursors
were observed in a few cases~\citep{minpf91}.  Between 1989 and 1994,
95 bursts were detected by the GRANAT/WATCH all-sky monitor in two
energy bands, 8--20~keV and 20--60~keV, and thirteen of them were
found to exhibit significant emission in the lower energy band before
and/or after the activity in the higher energy band~\citep{sstlb98}.
 
The Wide Field Camera on \sax~was used to detect 45 GRBs in the
1.5--26.1~keV band between July 1996 and February 2001\footnote{See
also J. Greiner's archive at {\tt
http://www.aip.de:8080/$\sim$jcg/grbgen.html}}.  Only one of those
bursts (GRB~980519) exhibited significant soft X-ray activity before
the onset of the GRB in gamma-rays~\citep{zhpf99}.  In addition to
providing broadband light curves for these GRBs~\citep{facmp00}, the
\sax~effort has revealed the existence of GRB ``afterglows'', the
fading emission sometimes seen after a GRB~\citep{cfhfz97}.
  
The general picture that has emerged is that X-ray light curves for
GRBs tend to track their gamma-ray counterparts.  Spectral evolution
may or may not be present.  Most often the times of X-ray peak
emission tend to lag behind the peaks at higher energies.  Bursts tend
to last longer in X-rays, but this is not true of every burst.  A
small fraction of bursts exhibit X-ray activity with no corresponding
emission at high energies.  This X-ray activity sometimes precedes the
gamma-ray burst proper and sometimes follows it.  In very rare cases,
it does both.  

\subsection{A Simplified Conventional Model\label{sec:grbmod}}

\citet{piran99} has reviewed the standard ``fireball'' model for GRBs
in great detail.  We present a brief summary of this model, to which
we shall refer when interpreting features of GRB light curves.  In
this model a large amount of energy ($\sim10^{51}$~ergs) is released
into a small volume to create a very hot, optically thick
fireball~\citep{goodm86}.  The fireball expands rapidly, and since it
contains only a very small amount of baryonic
matter~\citep{shpi90,kps99}, the expansion becomes highly
relativistic~\citep{blmck76}.  By the end of the acceleration phase,
all the available energy has been transferred to the bulk kinetic
flow~\citep{cavree78}.

The optical depth of an expanding fireball, $\tau$, can be related to
the bulk Lorentz factor of the expansion, $\Gamma$, and the minimum
observed timescale for variability during the burst, $\delta T$ ({\it
e.g.}~\citet{piran99}).  This relation hinges on the observation that
high-energy GRB spectra are non-thermal, with a photon index $\alpha$
($dN/dE \propto E^{-\alpha}$). In order for the observed spectrum to
be non-thermal, the optical depth must be less than unity.  The
derived relation is
\begin{equation}
\label{eq:tauyy}
\tau \sim \frac{2\times10^{15}}{\Gamma^{4+2\alpha}}\ E_{52}\ 
\left( \frac{\delta T}{10\ {\rm ms}} \right)^{-2} \lesssim 1,
\end{equation}
where $E_{52}$ is the total energy in the fireball in units of
$10^{52}$~ergs. This constraint demands a lower limit on the bulk Lorentz
factor, dependent on $\alpha$.  Early work found that the high energy spectral
index can vary from $\sim1.6$ to higher than 5, with no particular preferred
value~\citep{bmfsp93}.  A recent analysis of bright bursts from the Fourth
BATSE Catalog reports an asymmetric distribution of high-energy spectral
indices that peaks around 2.25 and extends beyond 4~\citep{pbmpp00}.  If
$\alpha$ is 2.25, and the other terms in Equation~\ref{eq:tauyy} are of order
unity, then $\Gamma$ must be greater than $\sim60$.

Once such a highly relativistic speed is reached, the ejecta coast quietly
until one of two kinds of shocks form.  When the ejecta sweep up mass from the
surrounding

\end{multicols}
\begin{deluxetable}{lcccc}
\tablecolumns{5} 
\tablewidth{0pc} 
\tablecaption{Burst Times and Locations\label{postab}}
\tablehead{
Date of GRB &
Time of GRB &
Confirming &
R.A. &
Decl. \\
(yymmdd) & 
(hh:mm:ss) &
Satellite\tablenotemark{a}  &
(J2000) &
(J2000) 
} 

\startdata
960416  & 04:09:00 & ubk   & 04h15m27s & $+77\arcdeg10\arcmin$ \\
960529  & 05:34:34 &  k    & 02h21m50s & $+83\arcdeg24\arcmin$ \\
960727  & 11:57:36 & uk    & 03h36m36s & $+27\arcdeg26\arcmin$ \\
961002  & 20:53:55 & uk    & 05h34m46s & $-16\arcdeg44\arcmin$ \\
961019  & 21:08:11 & ubk   & 22h49m00s & $-80\arcdeg08\arcmin$ \\
961029  & 19:05:10 & k     & 06h29m27s & $-41\arcdeg32\arcmin$ \\
961216  & 16:29:02 & bk    & \nodata   & \nodata               \\
961230  & 02:04:52 & u     & 20h36m45s & $-69\arcdeg06\arcmin$ \\
970815  & 12:07:04 & ubks  & 16h08m33s & $+81\arcdeg30\arcmin$ \\
970828  & 17:44:37 & ubki  & 18h08m23s & $+59\arcdeg19\arcmin$ \\
971024  & 11:33:32 & bk    & 18h25m00s & $+49\arcdeg27\arcmin$ \\
971214  & 23:20:41 & ubnks & 12h04m56s & $+64\arcdeg43\arcmin$ \\
980703  & 04:22:45 & ubk   & 23h59m04s & $+08\arcdeg33\arcmin$ \\
981220  & 21:52:21 & uks   & 03h43m38s & $+17\arcdeg13\arcmin$ \\
990308  & 05:15:07 & ubk   & 12h23m11s & $+06\arcdeg44\arcmin$ \\
000301C & 09:51:39 & unk   & 16h20m19s & $+29\arcdeg26\arcmin$ \\
\enddata
\tablenotetext{a}{u - \uly; b - BATSE; k - Konus; s - \sax; n - {\it NEAR}; i - {\it SROSS-C}}
\end{deluxetable}
\begin{multicols}{2}

\noindent
medium, they decelerate and convert the
bulk kinetic energy to random motion and radiation~\citep{mr92,rm92}.
This type of shock is referred to as ``external''.  If the central
engine of the GRB source is erratic and emits multiple shells of
ejecta at different speeds, ``internal'' shocks may form as these
shells overtake each other, and radiation will be
emitted~\citep{kps97}.  The fading afterglow observed after many GRBs
is believed to originate in the cooling of the material behind the
external shock front~\citep{mr97}, while the complex temporal
structure of the GRBs themselves has been attributed to multiple
internal shocks~\citep{npp92,fmn96}.  In both internal and external
shocks, synchrotron emission is expected to be the dominant source of
radiation~\citep{snp96,piran99}, although internal shocks are expected
to be more efficient~\citep{sp97,kps97,ks01}.

In a model of post-shock synchrotron radiation, the frequency of peak
emission ($\nu_m$) increases as the bulk Lorentz factor of the shock
($\Gamma_{\rm sh}$) to the fourth power~\citep{snp96,spn98,wg99}.  The
inverse of this relationship is given by
\begin{equation}
\label{eq:peaklimgam}
\Gamma_{\rm sh} = 14\ \epsilon_B^{-1/8} \epsilon_e^{-1/2} n_1^{-1/8} 
\left( \frac{p-1}{p-2} \right)^{1/4}
\left( \frac{h \nu_m}{1\ {\rm keV}} \right)^{1/4}
\end{equation}
(see, {\it e.g.},~\citet{piran99}, Equation~105), where $n_1$ is the
density of the external medium in cm$^{-3}$, $\epsilon_B$ and
$\epsilon_e$ define the fractional energy transferred to the
post-shock magnetic field and electron distribution, respectively, and
$p$ is the index of the number spectrum of the Lorentz factor
distribution of the shocked electrons ($N (> \Gamma_{\rm sh}) \propto
\Gamma_{\rm sh}^{-p}$).

In this paper we present X-ray light curves for the GRBs detected with
the All-Sky Monitor (ASM) on the {\it Rossi X-ray Timing Explorer}
({\it \rxte}) and interpret them in the context of this synchrotron
shock model.  These light curves are for the GRBs described
in~\citet{sblr99}, to which we add those for GRB~990308 and
GRB~000301C.  We also present a light curve for GRB~961216, which was
detected but could not be localized accurately.  All these events have
been confirmed as GRBs via their detection at higher energies by other
instruments in the Interplanetary Network
(Table~\ref{postab})\footnote{A list of burst detections by the IPN is
maintained by K. Hurley at {\tt http://ssl.berkeley.edu/ipn3/}.}.
Satellites participating in the IPN during the time interval covered
by this work include {\it CGRO}, \wind, {\it SROSS-C}, {\it NEAR},
\sax, and \uly.  Section~\ref{sec:lcdat} describes the data and our
analysis techniques.  Section~\ref{sec:lcs} presents the observations
and the GRB light curves.  We compare the ASM light curves with the
associated high-energy light curves recorded by detectors on other
satellites.  Section~\ref{sec:lcsum} summarizes both the common
features and the striking differences among these light curves and
discusses them in the context of the model.

\section{INSTRUMENTATION AND ANALYSIS\label{sec:lcdat}}

The ASM consists of three Scanning Shadow Cameras (SSCs) mounted on a motorized
rotation drive~\citep{lbcjm96}.  The assembly holding the three SSCs is
generally held stationary for a 90-s ``dwell''.  The drive rotates the SSCs
through $6\arcdeg$ between dwells, except when it is necessary to rewind the
assembly.  Each SSC contains a proportional counter with eight resistive anodes
and views

\end{multicols}
\begin{deluxetable}{lrrr}
\tablecolumns{4} 
\tablewidth{0pc} 
\tablecaption{Burst Fluxes and Fluences\label{flutab}}
\tablehead{
Date of GRB &
1.5--12 keV Fluence\tablenotemark{a} & 
50--200 keV Fluence\tablenotemark{b} &
1.5--12 keV Peak Flux\tablenotemark{a} \\
(yymmdd) & 
($10^{-7}$ ergs cm$^{-2}$) &
($10^{-7}$ ergs cm$^{-2}$) &
($10^{-8}$ ergs cm$^{-2}$ s$^{-1}$)
} 

\startdata
960416  & $  6.0\pm0.3$ & 8.0     & $7.8\pm0.8$   \\
960529  & $>17.5\pm0.6$ & 58      & $2.2\pm0.2$   \\
960727  & $  9.5\pm0.5$ & 84      & $6.0\pm0.5$   \\
961002  & $  9.2\pm0.5$ & 35      & $7.1\pm0.6$   \\
961019  & $  4.6\pm0.6$ & 14      & $3.1\pm0.7$   \\
961029  & $ >3.3\pm0.4$ & 20      & $>3.2\pm0.5$  \\
961216  & \nodata       & 12      & \nodata       \\
961230  & $  1.5\pm0.3$ & \nodata & $0.65\pm0.08$ \\
970815  & $>33.3\pm0.8$ & 80      & $6.1\pm0.5$   \\
970828  & $>14.9\pm0.6$ & 280     & $4.0\pm0.5$   \\
971024  & $ >1.1\pm0.3$ & 40      & $0.4\pm0.1$   \\
971214  & $ 3.4\pm0.3$  & 36      & $0.9\pm0.2$   \\
980703  & $>18.3\pm0.8$ & 72      & $3.2\pm0.3$   \\
981220  & $ 12.6\pm0.5$ & 52      & $8.2\pm0.9$   \\
990308  & $ 5.8\pm0.4$  & 18      & $1.3\pm0.3$   \\
000301C & $ 3.8\pm0.5$  & 16      & $4.7\pm0.9$   \\
\enddata
\tablenotetext{a}{75.5 c/s in ASM units $\equiv$ 1~Crab $\equiv 2.8\times 10^{-8}$ erg~cm$^{-2}$~s$^{-1}$}
\tablenotetext{b}{Derived from Konus observations}
\end{deluxetable}
\begin{multicols}{2}

\setcounter{figure}{0} 
\refstepcounter{figure} 
\PSbox{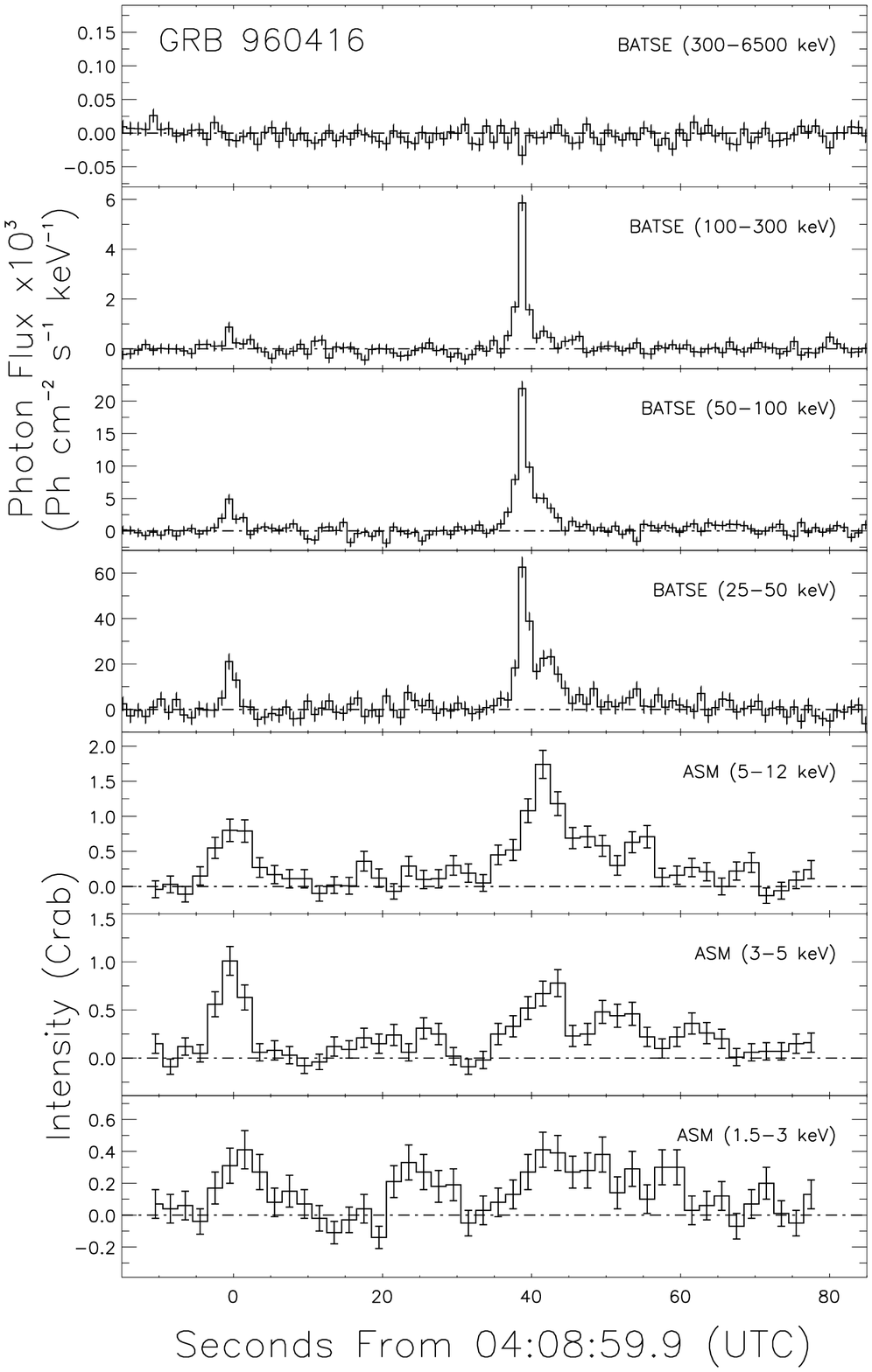 hoffset=-22 voffset=-13 hscale=58 vscale=58}{8.8cm}{14.35cm}
{\\\\\small Fig. 1 -- Time-series data for GRB~960416 in six energy
channels as recorded by both the ASM (2-s bins; 1.5--12~keV) and BATSE
(1-s bins; 25--6500~keV).  The ASM light curve is the weighted average
of measurements by both SSC~1 and SSC~2.  Where possible, the ASM
light curves presented in these figures have been converted into Crab
flux units by the method described in
Section~\protect{\ref{sec:lcdat}}.}
\label{fig:lc960416}

\pfg
\noindent
a $12\arcdeg \times 110\arcdeg$ (FWZI) field through a mask perforated
with pseudo-randomly spaced slits.  The long axes of the slits run
perpendicular to the anodes.  The net effective area for a source at
the center of the FOV is $\sim30$~cm$^2$.

During each dwell, the positions of incident photons are tabulated in
histograms for each anode for each of three pulse height channels:
1.5--3, 3--5, and 5--12~keV, denoted A, B, and C, respectively.  The
logical union of these three bands is referred to as the ``sum band'',
or ``S''.  In this position histogram mode, the arrival time of each
photon is not preserved.  A histogram contains information on the
pattern of illumination of the detector through the slit mask from
each discrete X-ray source in the field of view (FOV), as well as
contributions from the diffuse X-ray background and local particle
events.

For each of the three energy channels, we carry out a fit of these
data to obtain the strength of each known X-ray

\refstepcounter{figure} 
\PSbox{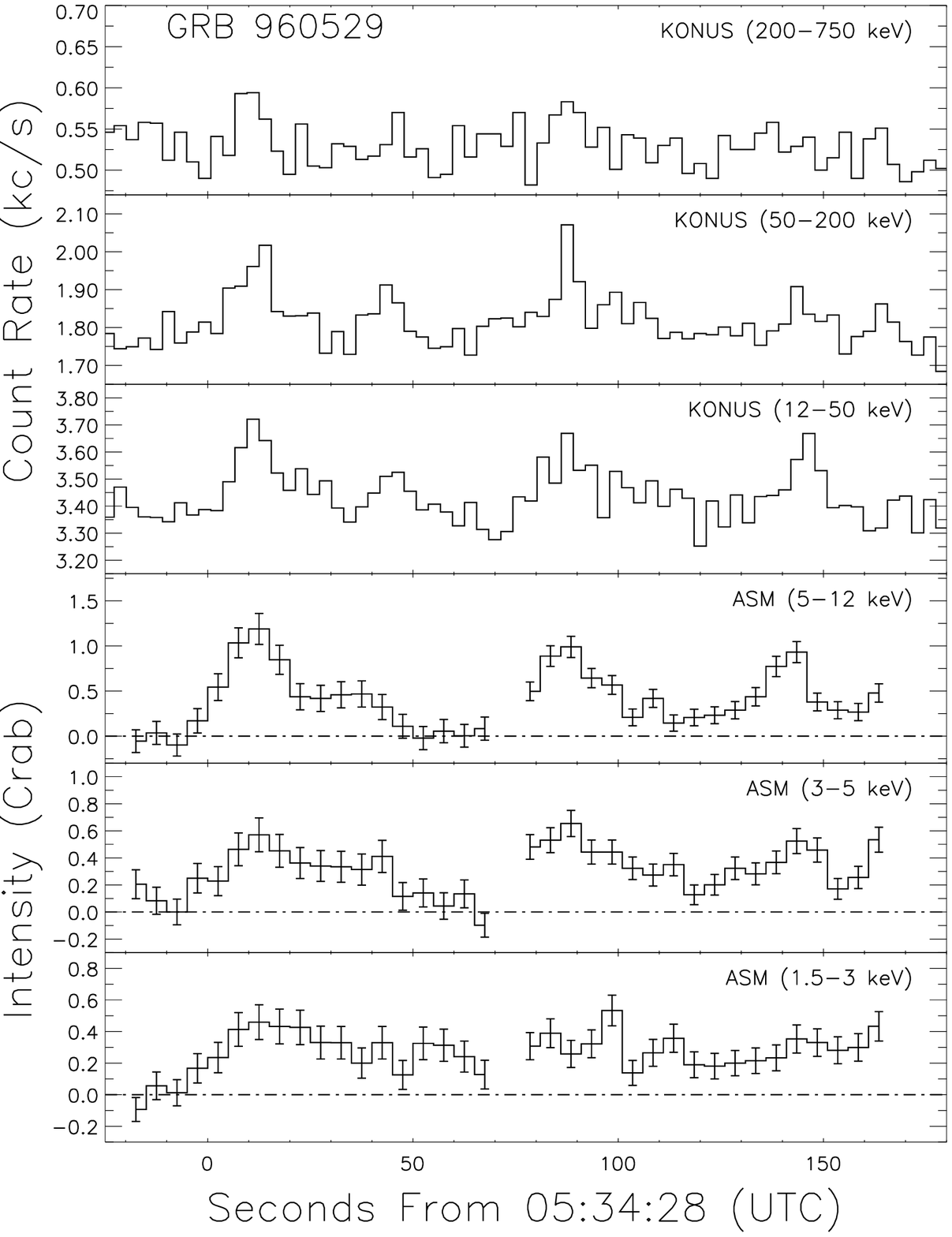 hoffset=-15 voffset=-10 hscale=56 vscale=56}{8.8cm}{11.9cm}
{\\\\\small Fig. 2 -- Time-series data for ASM and Konus observations
of GRB~960529.  The ASM light curve is the weighted average of
measurements by both SSC~1 and SSC~2, in 5-s bins.  A gap indicates
the 6~s interval between dwells when the ASM assembly was in motion.
The Konus count rates are from the ``waiting-mode'' data in 3-s bins.}
\label{fig:lc960529}

\pfg
\noindent
source in counts per $\sim30$~cm$^2$ for the entire 90-s dwell. Division by the
exposure time and the application of a multiplicative correction factor yields
the intensity in ``ASM units''.  The time-dependent correction factor ($\equiv
a$) is empirically determined such that the corrected intensity of the Crab
Nebula is the count rate that would be obtained for the source had it been at
the center of the FOV of SSC~1 at a fiducial time near the start of the
mission.  The ASM units for 1.0 Crab are 26.8, 23.3, 25.4, and 75.5 counts
s$^{-1}$ for the A, B, C, and Sum channels respectively.  Table~\ref{flutab}
shows the burst fluences and peak fluxes as measured by the ASM in the Sum
band.  ASM units are converted to physical units by assuming a Crab-like
spectrum~\citep{sew78} in the 1.5--12~keV band, which yields an integrated flux
of $2.8\times10^{-8}$~erg~cm$^{-2}$~s$^{-1}$.

A second data mode records the total number of 1.5--12~keV events
detected from all sources in the FOV of each SSC.  These data are
recorded in 0.125-s bins in each of the same three energy channels as
the position histogram mode.  In this ``multiple time series'' (MTS)
data mode, the physical location in the detector of each incident
photon is not preserved.

In combination, these two data modes can yield a light curve for a
highly variable source like a GRB.  The 

\refstepcounter{figure} 
\PSbox{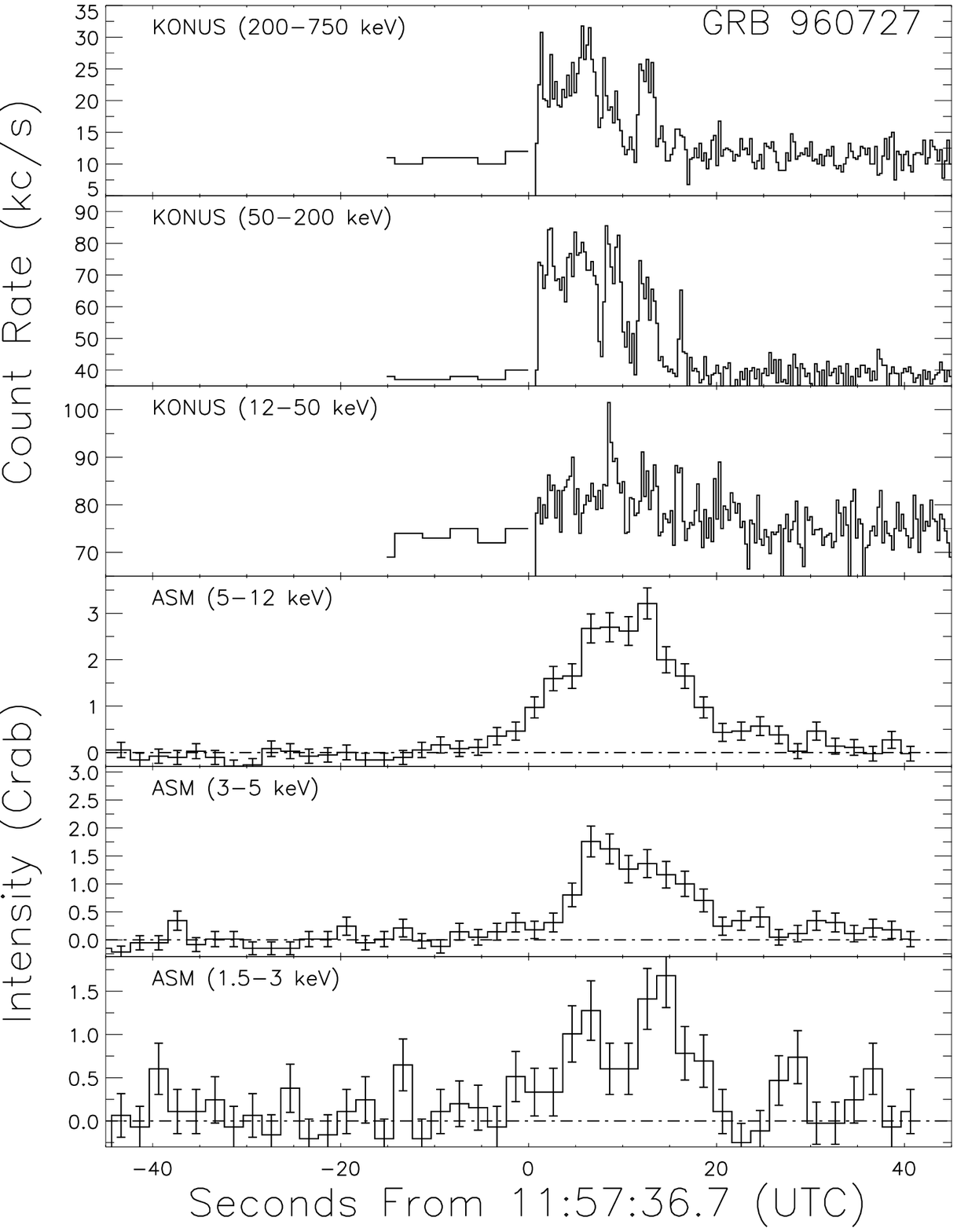 hoffset=-10 voffset=-12 hscale=55 vscale=55}{8.8cm}{11.5cm}
{\\\\\small Fig. 3 -- ASM time-series data from SSC~2 for GRB~960727
(2-s bins) in three energy bands, as well as the corresponding count
rate histories from Konus (0.25-s bins after the trigger) in three
energy bands.}
\label{fig:lc960727}

\pfg
\noindent
challenge these data present is to properly set the background level in the MTS
data.  We take advantage of the fact that no known variable X-ray sources were
in the FOV during the observations of any of the GRBs presented here and assume
that the count rates from all sources other than the GRB do not vary during the
dwell. Should the FOV contain variable sources other than the GRB, or should
the background rate vary, the method described here would be invalid.  The
total number of counts from the GRB source detected during one dwell of one SSC
is obtained from the standard fit to the position histogram data.  We then
define the effective background level (for the GRB) in the MTS count rate data
such that the total number of counts above that level for the appropriate 90-s
interval is equal to the number of counts inferred from the position histogram
data to come from the GRB source.  The number of MTS counts from the GRB
recorded in a given time bin can therefore be estimated as:
\begin{equation}
c_j = n_j - t_j\ \left(\frac{N}{T} - \frac{fR}{a}\right),
\label{eq:bgcounts}
\end{equation}
where the term in parentheses is the inferred background rate.  $N$ is
the total number of MTS counts detected in a given SSC energy band
during an observation with total exposure time $T$, $R$ is the
time-averaged GRB source intensity in ASM units derived from the fit
to the position-histogram data, $f$ is the transmission fraction for
the location of the GRB in the FOV, $a$ is the time-dependent

\refstepcounter{figure} 
\PSbox{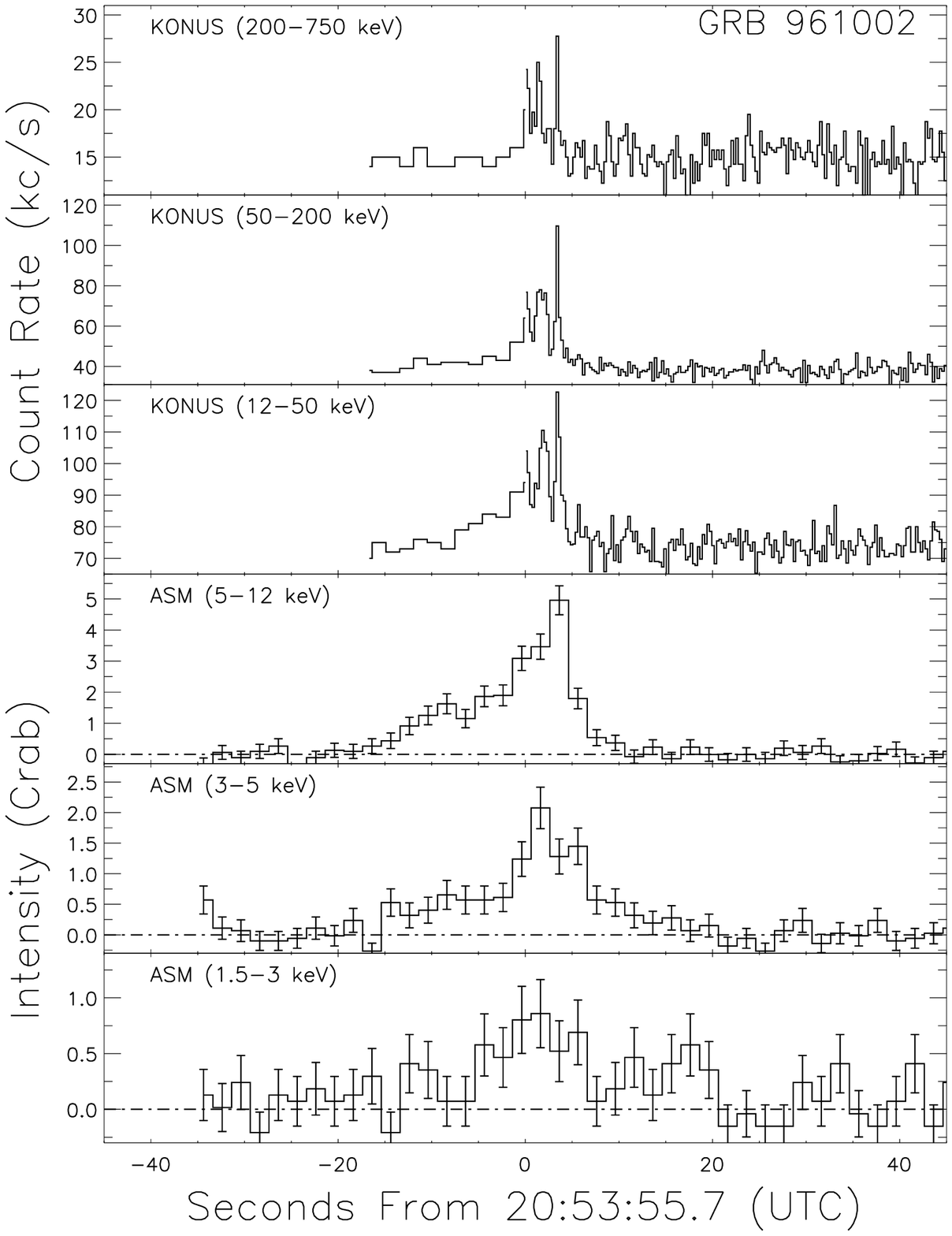 hoffset=-10 voffset=-5 hscale=55 vscale=55}{8.8cm}{11.7cm}
{\\\\\small Fig. 4 -- ASM time-series data from SSC~2 for GRB~961002
(2-s bins) in three energy bands, as well as the corresponding count
rate histories from Konus (0.25-s bins after the trigger) in three
energy bands.}
\label{fig:lc961002}

\pfg
\noindent
correction factor, $t_j$ is the time bin size in seconds, and $n_j$ is
the number of counts in the $j$th time bin.  One can then convert
$c_j$ into ASM units through multiplication by $a/f$.  For additional
details see~\citet{smithe99}.  

We compare the resulting ASM light curves with available
contemporaneous BATSE, \uly, Konus, or \sax~GRBM count rates.  The
BATSE count rates shown here are extracted from the {\tt discsc} mode
FITS files and converted to photon flux rates according to the method
described in~\citet{ppbmk94,pmpbp96,ppbpm97}.  This method combines a
photon spectral model with a coarse spectral inversion technique.  A
polynomial defined in Log photon energy -- Log photon flux across the
four broad energy bins is used as the spectrum to construct the
detector response matrix (DRM) prior to the application of a direct
spectral inversion technique. The first inversion yields a preliminary
coarse photon spectrum that is used to re-evaluate the coefficients of
the GRB's spectral polynomial. This updated polynomial is used to
build a second DRM employed in a second application of the spectral
inversion technique that produces the final photon flux data.  The
data produced with this method are also used to generate the BATSE GRB
peak flux data for the BATSE burst catalogs.  These data are presented
in four energy channels with nominal energy ranges of 25--50, 50--100,
100--300, and 300--6500~keV.  In one case (GRB~970828),

\refstepcounter{figure} 
\PSbox{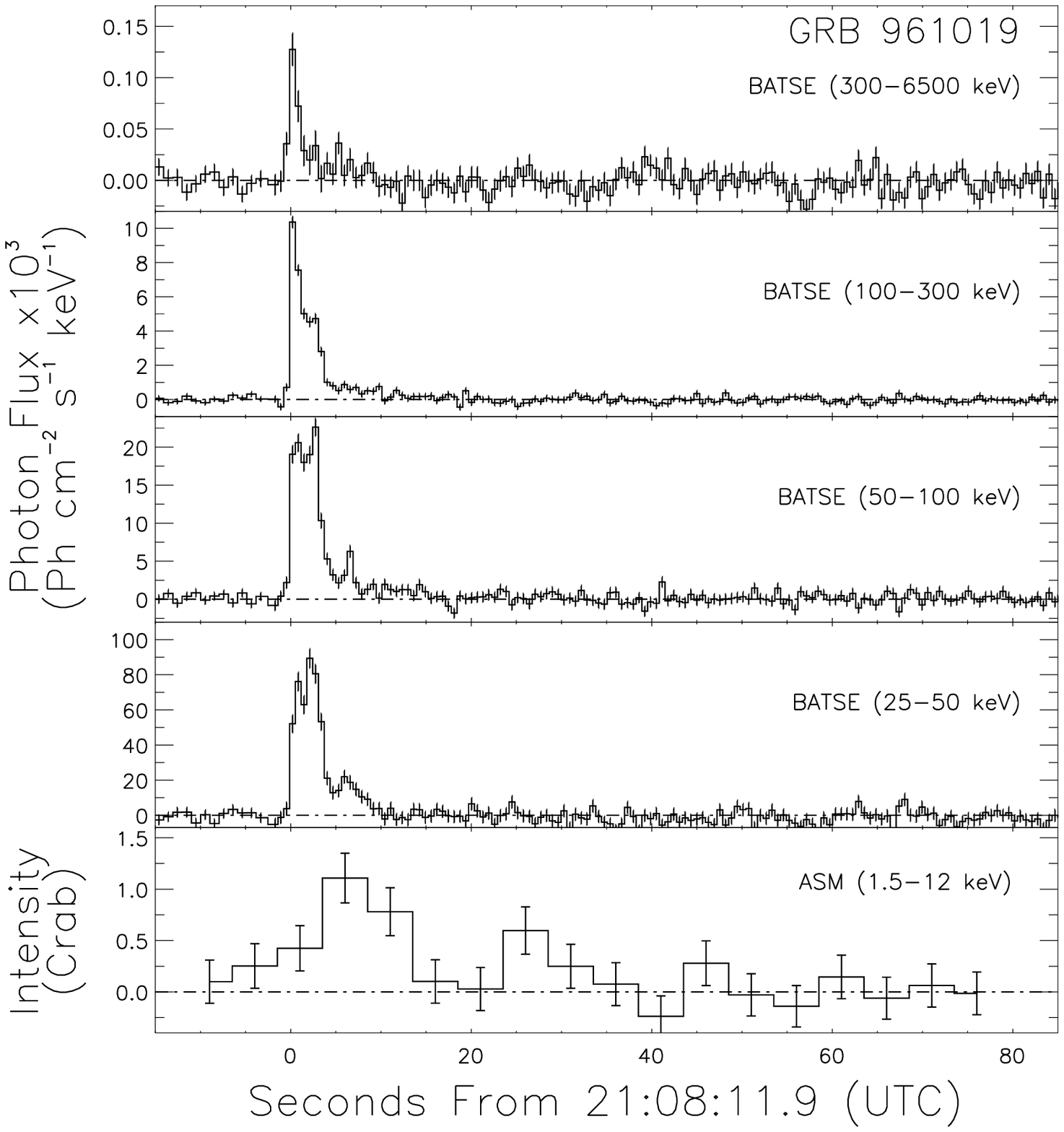 hoffset=-15 voffset=-10 hscale=53 vscale=53}{8.8cm}{9.7cm}
{\\\\\small Fig. 5 -- Time-series data from SSC~2 for GRB~961019 in both the
ASM (5-s bins; 1.5--12~keV) and BATSE (0.64-s bins; 25--6500~keV).}
\label{fig:lc961019}

\pfg 
\noindent
the event occurred during a telemetry gap, so no {\tt discsc} data were
available.  In this case, we used the 16-channel {\tt MER} data and combined
channels to approximate the energy ranges provided by the {\tt discsc} mode.
The conversion technique described above could not be applied to these data, so
we present here the count rates without background subtraction.

Where BATSE data are not available, we compare the ASM light curves
with the count rates from other high-energy instruments.  The hard
X-ray detectors for the solar X-ray/cosmic gamma-ray burst experiment
on \uly~record 25--150~keV photons in time bins of size between
0.25--2~s, depending on the telemetry rate available, switching to
0.03125-s bins on a trigger~\citep{hsabc92}.  When the \sax~GRBM
registers a burst, the 40--700~keV count rate is stored in bins with a
minimum size of 7.8~ms from 8 s before to 98 s after the trigger time;
all on-board data is sent down once every $\sim90$-min orbit for
operators to examine~\citep{fcffn97,ffcda97}.  The Konus instrument on
\wind~records data continuously in a ``waiting mode'', with bins of
2.9 s, but when a trigger is activated, counts are recorded in higher
resolution for a certain amount of time after the
trigger~\citep{afgim95}.  The triggered data presented in this paper
have been binned to 256~ms.  Since Konus detected all but one of the
GRBs presented here, the burst fluence as measured in the Konus
50--200~keV channel is presented in Table~\ref{flutab} to provide a
consistent comparison with the 1.5--12~keV fluence as measured by the
ASM.  The times of the bins in the high energy light curves obtained
from distant instruments such as \uly~or Konus have been shifted to
account for the propagation times.  Light travel times between
instruments in near Earth orbit are negligible.

\refstepcounter{figure} 
\PSbox{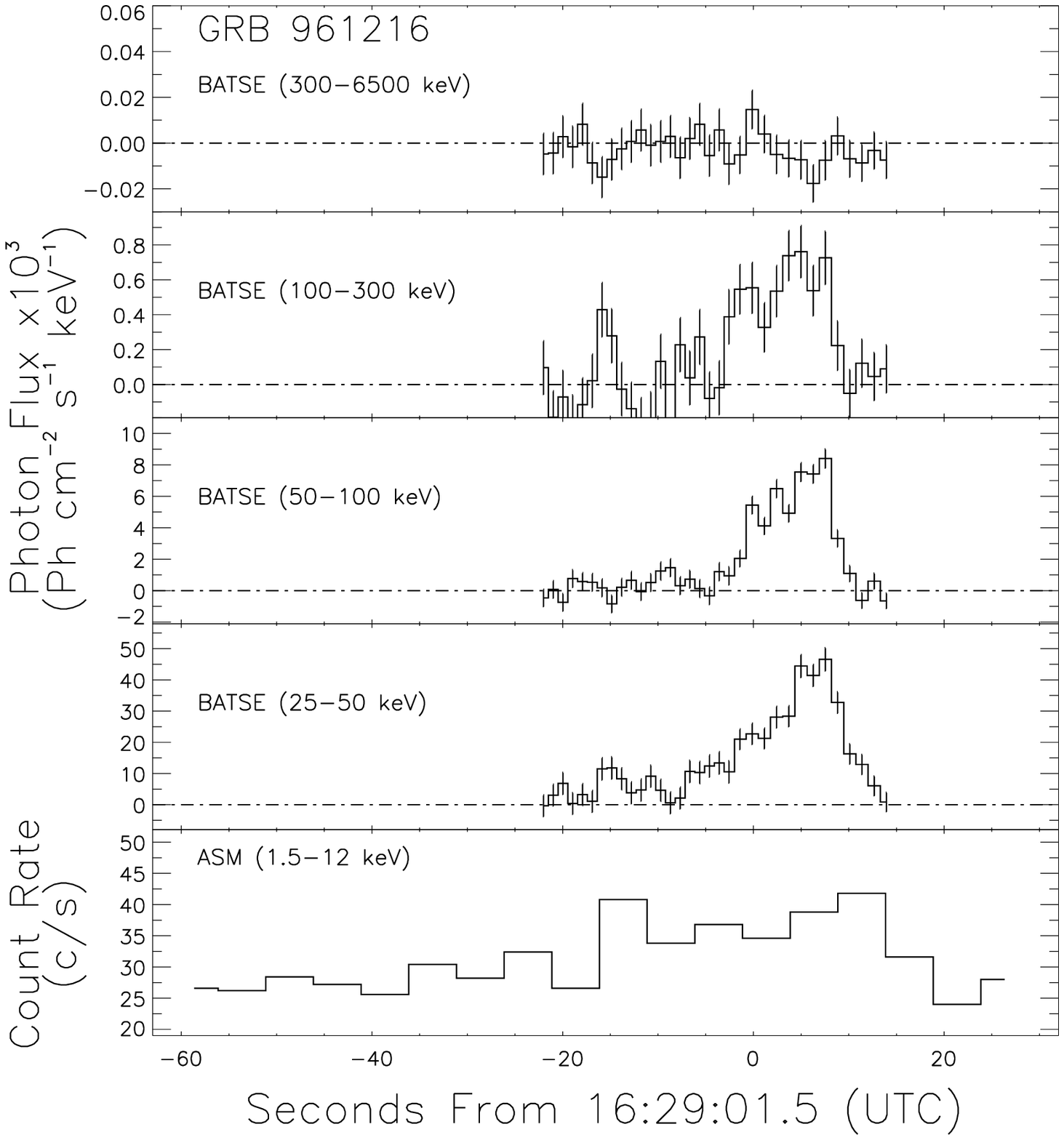 hoffset=-15 voffset=0 hscale=53 vscale=53}{8.8cm}{10.0cm}
{\\\\\small Fig. 6 -- Time-series data from SSC~2 for GRB~961216 in both the
ASM (5-s bins; 1.5--12~keV) and BATSE (1-s bins; 25--6500~keV).  No
background subtraction has been performed on the ASM data.}
\label{fig:lc961216}

\pfg

\section{OBSERVATIONS\label{sec:lcs}}

\citet{sblr99} describe a search through 1.5~years of archived ASM
time series data, as well as an ongoing program to scrutinize the ASM
data on a near real-time basis for evidence of X-ray counterparts to
GRBs.  The near real-time program used three methods in conjunction: a
low-threshold search for unidentified X-ray sources in ASM data around
the times of bursts detected by BATSE or other IPN instruments, a
search for variability in ASM MTS data, and examination of all ASM
position histogram data for evidence of uncatalogued transient
sources.

The efficiency of these searches for the detection of X-rays from GRBs
depended on a number of factors.  For any particular dwell, these
included (1)~the 2--12~keV peak flux and duration of the GRB, (2)~the
position of the GRB source within the field of view, (3)~whether any
other strong sources were in the field of view, and (4)~the background
level and variability, which often were affected because of passage
through a region of high flux of energetic charged particles or
because of contamination by scattered solar X-rays.

The variability search measured the deviations from a linear fit to
the count rate over the given dwell.  Thus, GRBs in which the X-ray
flux changed slowly within the 90-s duration of a dwell were difficult
to find.  For example, if the ASM were to scan onto a slowly fading
GRB in progress, the observation would not produce a substantial
deviation from a linear trend, and would not be flagged in the
variability search.  We also ignored dwells with known strong sources
in the FOV, because even if variability were 

\refstepcounter{figure} 
\PSbox{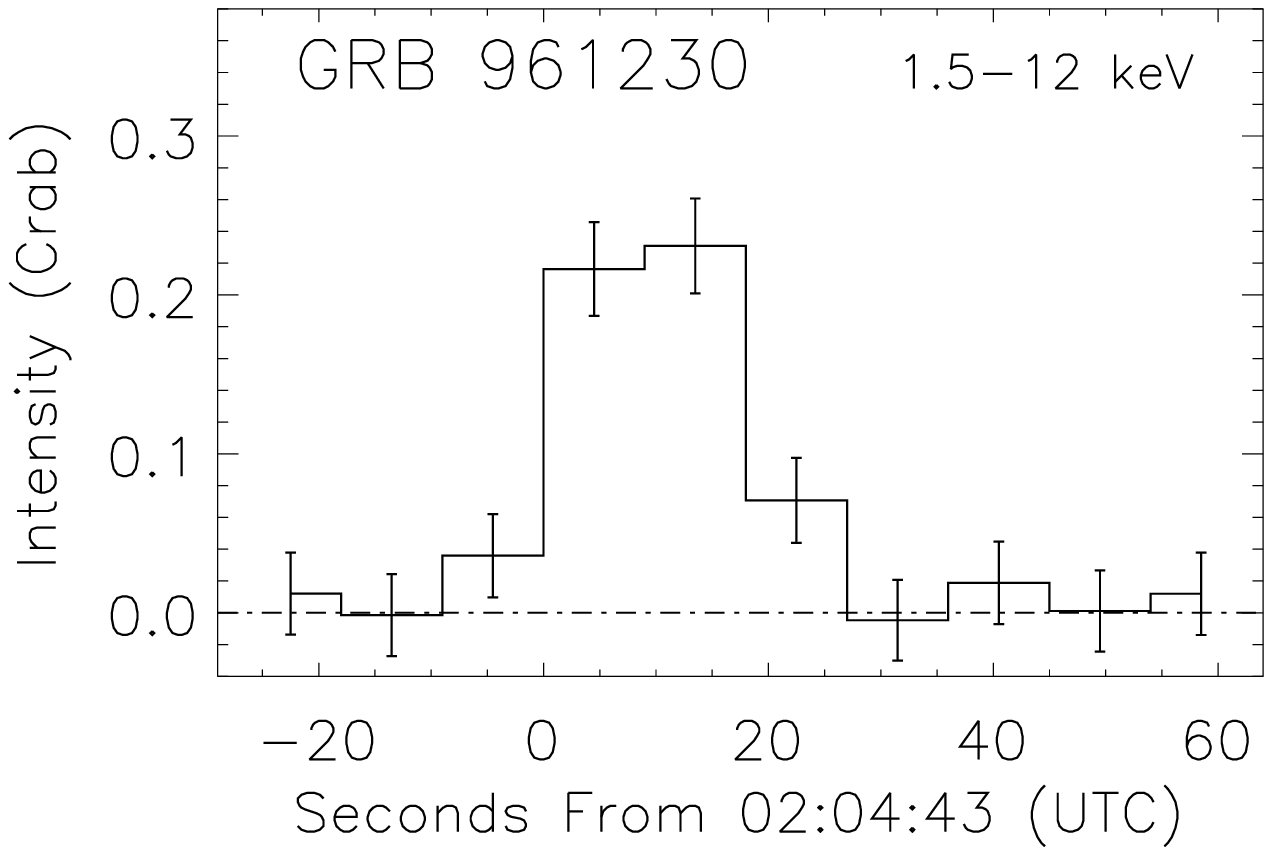 hoffset=-20 voffset=-10 hscale=63 vscale=63}{8.8cm}{5.5cm}
{\\\\\small Fig. 7 -- Time-series data for GRB~961230 (9-s bins;
1.5--12~keV), displaying the weighted average of measurements by both
SSC~1 and SSC~2.}
\label{fig:lc961230}

\pfg
\noindent
apparent, it would be difficult to determine the source of that variability.

We know of no bias against the detection of short GRBs in these
searches, except that the shorter the duration of the X-ray event, the
higher the peak flux must be to yield sufficient counts for the GRB to
be detected in the position-histogram data.  We found no ASM
detections of any GRB counterparts which lasted less than a few
seconds.  Our results strongly suggest that short GRBs do not have
X-ray fluences as high as those of longer GRBs.

As noted above, we searched for ASM counterparts of GRB events
detected by BATSE and other high-energy instruments.  We found no GRB
event which is known through these other instruments to have occurred
in the FOV of the ASM but was not detected.  However, the
localizations for most GRBs detected by higher energy instruments
could not be constrained to lie completely within the ASM FOV.

Here we present, in chronological order, the ASM light curves for
15 of the GRBs discovered in these searches.

GRB~960416 (Fig.~\ref{fig:lc960416}) was observed in both SSCs~1 and~2
during a single dwell.  The BATSE light curve shows two distinct peaks
$\sim40$~s apart.  The two peaks are also seen in the ASM data, but
the ASM reveals a third, remarkably soft, peak between them, as well
as an extended tail for the final peak beyond the end of the BATSE
event.

GRB 960529 (Fig.~\ref{fig:lc960529}) exhibits three hard peaks in the
ASM time-series data from two SSCs over the course of two successive
dwells.  Although no Konus trigger was explicitly activated, the
waiting-mode data show the multi-peak structure of this burst
out to 200~keV.  The Konus light curve contains four major peaks,
such that the extended tail of the first peak in the ASM light curve
is resolved into two distinct peaks at higher energies.
 
GRB~960727 and GRB~961002 (Figs.~\ref{fig:lc960727}
and~\ref{fig:lc961002}) were each detected only in SSC~2.  Each lasted
about 30~s, and each showed a singly-peaked soft X-ray light curve
without strongly significant structure on time scales down to
$\sim1$~s.  Neither burst was detected by BATSE, but each was detected
by Konus and the GRB detector on \uly.  The high-energy light curves
reveal rich temporal structure.  The ASM light curves do not show
corresponding structure, but this is, at least in part, due to reduced
sensitivity to variability on subsecond time scales.  Each event
seems to conclude with a weak, extended X-ray tail, absent at high
energies, that lasts for 10 or 20~s.  The accuracy of the background
estimation before the burst supports the existence of a post-burst
X-ray excess, but the extended tail is too weak in the three ASM
sub-bands to make a useful measurement of the energy-dependence of the
decay rates.

GRB~961019 (Fig.~\ref{fig:lc961019}) was detected in a single
observation of SSC~2 and was also observed with BATSE.  The BATSE
light curve shows three sub-peaks.  The GRB was only
$1\arcdeg.5\pm0.1$ from the edge of the SSC FOV, so that only
24\% of the detector surface was exposed to the source.  The X-ray
peak emission may be delayed relative to the gamma-ray maximum by
5--10~s.

GRB~961029 (not shown) was detected as a dramatic rise in count rate
only a few seconds from the end of a dwell.  During this dwell, the
source was located only $2\arcdeg.0\pm0.2$~\citep{sblr99} from the
edge of the FOV of SSC~2.  As SSC~2 was rotated after the end of the
dwell, the field of view moved off the direction to the GRB source and
the signal was lost.  Konus reported a burst detection at 19:05:10
(UTC), which is during the rise of the ASM event.  No other
high-energy GRB detector observed this event.

GRB~961216 (Fig.~\ref{fig:lc961216}) was detected by a single SSC.
The location of GRB~961216 lay only $\sim1\arcdeg$ from the edge of
the FOV, which is outside the region for which our
position-determining ability is well-calibrated~\citep{sblr99}.  We
therefore show the total 1.5--12~keV count rate as measured by the
entire SSC without background subtraction.  

GRB~961230 (Fig.~\ref{fig:lc961230}) was a weak burst that was
detected in both SSC~1 and SSC~2 during the same dwell.  The X-ray
flux lasted about 25~s and reached a peak of $0.23\pm0.03$~Crab.  This
burst was detected by the GRB detector on \uly~at 02:04:52 (UTC) but
not by any other instrument.

GRB~970815 (Fig.~\ref{fig:lc970815}) exhibited multiple peaks over an
interval of several minutes.  SSC~2 scanned onto the source during the
decay from an initial peak.  The decay timescale from this peak is
clearly longer at lower energies, across all seven available energy
channels, indicative of the spectral softening common to the early
phases of GRB decay curves.  A second peak began 70~s into the dwell,
and a third peak $\sim50$~s after that, during the following dwell.
The 90-s averaged flux from the GRB in the third dwell, during which
the burst source was $\sim0\arcdeg.75$ from the edge of the FOV of
SSC~1, is $30\pm20$~mCrab (1.5--12~keV).  Figure~\ref{fig:lc970815_c}
shows the second peak on an expanded scale.

It is striking that the third peak, the strongest in the ASM light
curve, barely registers in BATSE's two lowest energy channels
(25--100~keV).  The X-ray spectrum of the burst evolves rapidly during
this peak, as indicated by a distinct soft lag of $\sim8$~s in the
times of peak burst flux between the high and low energy channels of
the ASM.  The ASM spectrum during this peak is clearly different from
that during the second peak.  We assume a simple power law spectrum
($N\propto E^{-s}$) without absorption in the 1.5--12~keV band and
find the index $s$ for the third peak is $1.8\pm0.1$ while the second
peak is harder, with an index of $1.2\pm0.3$.  This change may also be
described by a shift in the break frequency of the canonical, broken
power-law, 

\refstepcounter{figure} 
\PSbox{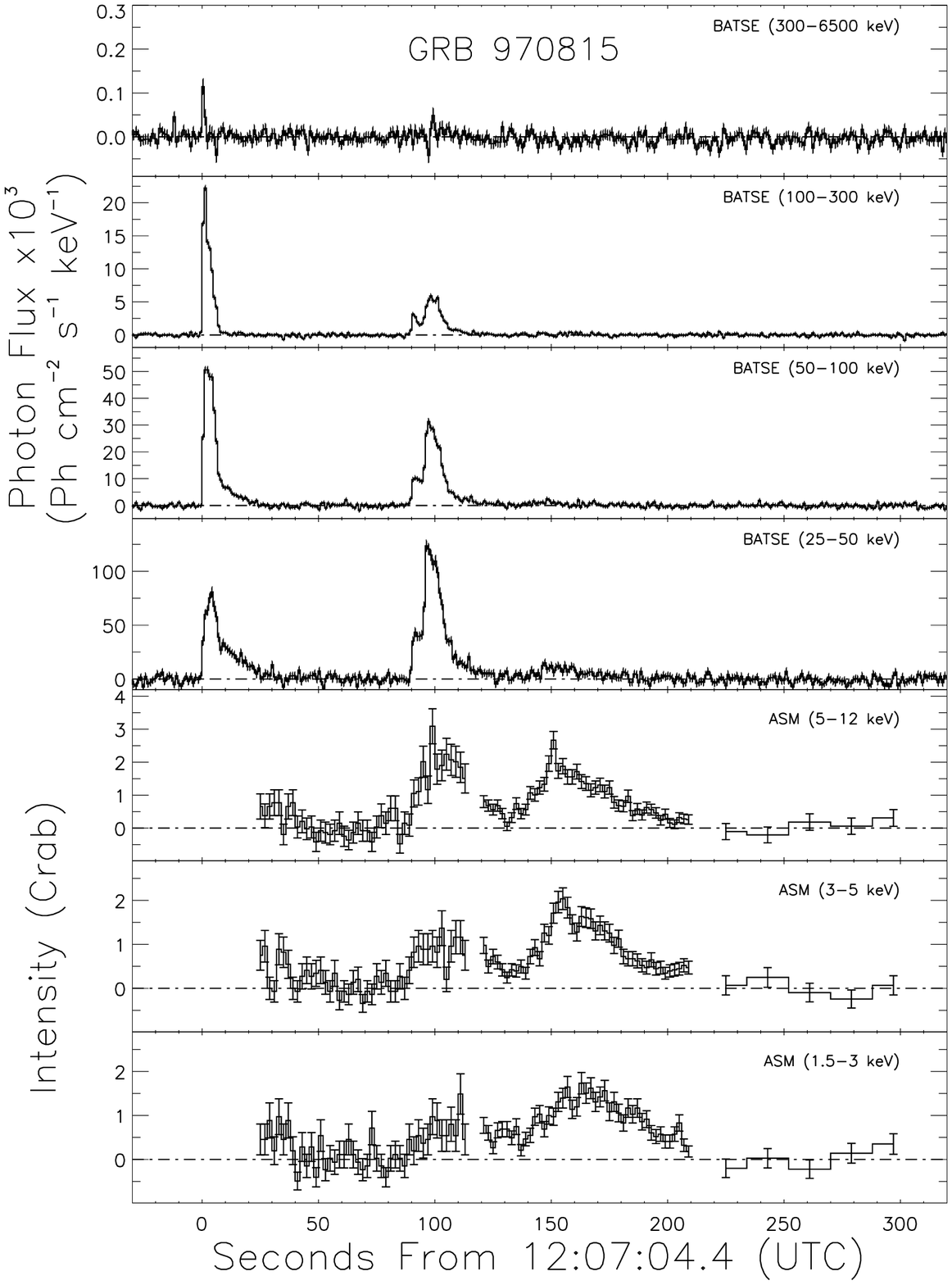 hoffset=-15 voffset=-22 hscale=52 vscale=52}{8.8cm}{11.5cm}
{\\\\\small Fig. 8 -- Light curves for GRB~970815 as measured by both the ASM
and BATSE.  The ASM scanned SSC~2 onto the GRB location during the decay of the
first peak.  Gaps in the ASM light curve indicate the 6~s intervals between
dwells when the ASM assembly was in motion.  The first two dwells are graphed
in 2-s bins.  The second dwell is represented here by the weighted average of
SSCs~1 and~2.  The GRB is very dim during the third dwell, and although the
data from SSC~1 have been binned into 9-s bins, the flux in each bin is
consistent with zero.  The 90-s averaged flux from the GRB in this dwell is
$30\pm20$~mCrab (1.5--12~keV).  The BATSE light curve is presented in 1-s bins
(25--6500~keV).}
\label{fig:lc970815}

\pfg
\noindent
``Band'' spectrum.  These indices were determined from a
5-s interval beginning 99~s after the BATSE trigger time and a 2-s
interval beginning at 154~s.

GRB~970828 (Fig.~\ref{fig:lc970828}) was a bright burst detected in
both SSC~1 and 2~\citep{rwsl97}.  The burst onset was observed midway
through a 90-s dwell.  Its FOV location was such that it was only
observed by SSC~1.  The ASM drive assembly rotated the SSCs between
dwells while the burst was still active, and the new aspect placed the
burst's FOV location just $0\arcdeg.5$ inside the edge of the FOV of
SSC~2.  Despite the reduction in effective area, the counting rate
during the second dwell yields a clear detection of the GRB in at
least two of the three energy channels.  A second rotation brought the
source to the center of the FOV of SSC~2, in time to witness the final
decay of the event.  Spectral softening during this decay is apparent.

At the onset, the X-ray flux climbs more slowly than the gamma-ray flux, as
shown in Figure~\ref{fig:lc970828_c}, but the 5--12~keV

\refstepcounter{figure} 
\PSbox{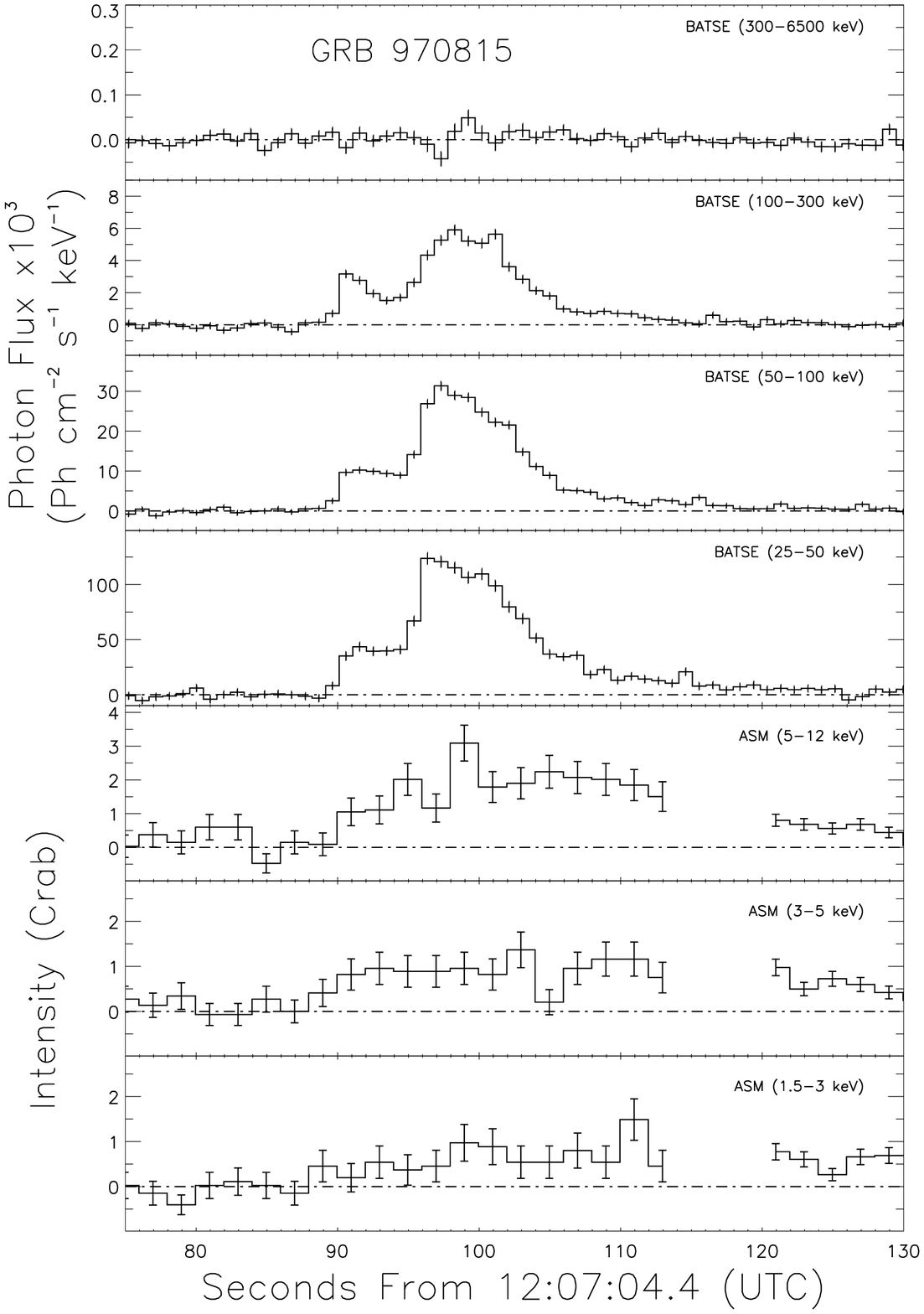 hoffset=-18 voffset=-15 hscale=53 vscale=53}{8.8cm}{12.7cm}
{\\\\\small Fig. 9 -- Light curves for GRB~970815
(Fig.~\protect{\ref{fig:lc970815}}) for the interval between 70 and 130~s after
the BATSE trigger time, showing the second peak.  There is a six-second gap at
114~s in the lower panels because the ASM assembly was in motion.}
\label{fig:lc970815_c}

\pfg
\noindent
structure appears to echo the gamma rays.  The time of peak emission may lag at
lower X-ray energies.  After 40~s, the high-energy flux drops below detectable
levels, then subsequently flashes through at least five further peaks.
Emission is detected by the ASM during the time interval around these peaks,
but the counting statistics are too weak to allow the individual peaks to be
resolved, even if they are present.  The X-ray emission appears to last long
after the cessation of gamma-ray activity.

GRB~971024 (Fig.~\ref{fig:lc971024}) was an extremely weak burst in
all energy bands.  It was detected in both SSC~1 and SSC~2, but the
latter yielded only an average flux over 90~s.  Large systematic
uncertainties in the estimation of the source flux stem from
relatively large uncertainties in the source position~\citep{sblr99}.
The BATSE light curves show obvious spectral softening over the
$\sim110$~s of the decay, but the ASM light curve is too short and too
weak for fruitful comparison.

GRB~971214 (Fig.~\ref{fig:lc971214}) appeared as a moderate-intensity
single-peak event, lasting $\sim40$~s, observed with SSC~3 during a single
dwell.  Limited statistics do not allow us to probe the soft X-ray light curve
for counterparts to the complex structures in the BATSE hard X-ray data, {\it
e.g.} the sharp spike at 32~s.  As with GRB~970828, the 

\refstepcounter{figure} 
\PSbox{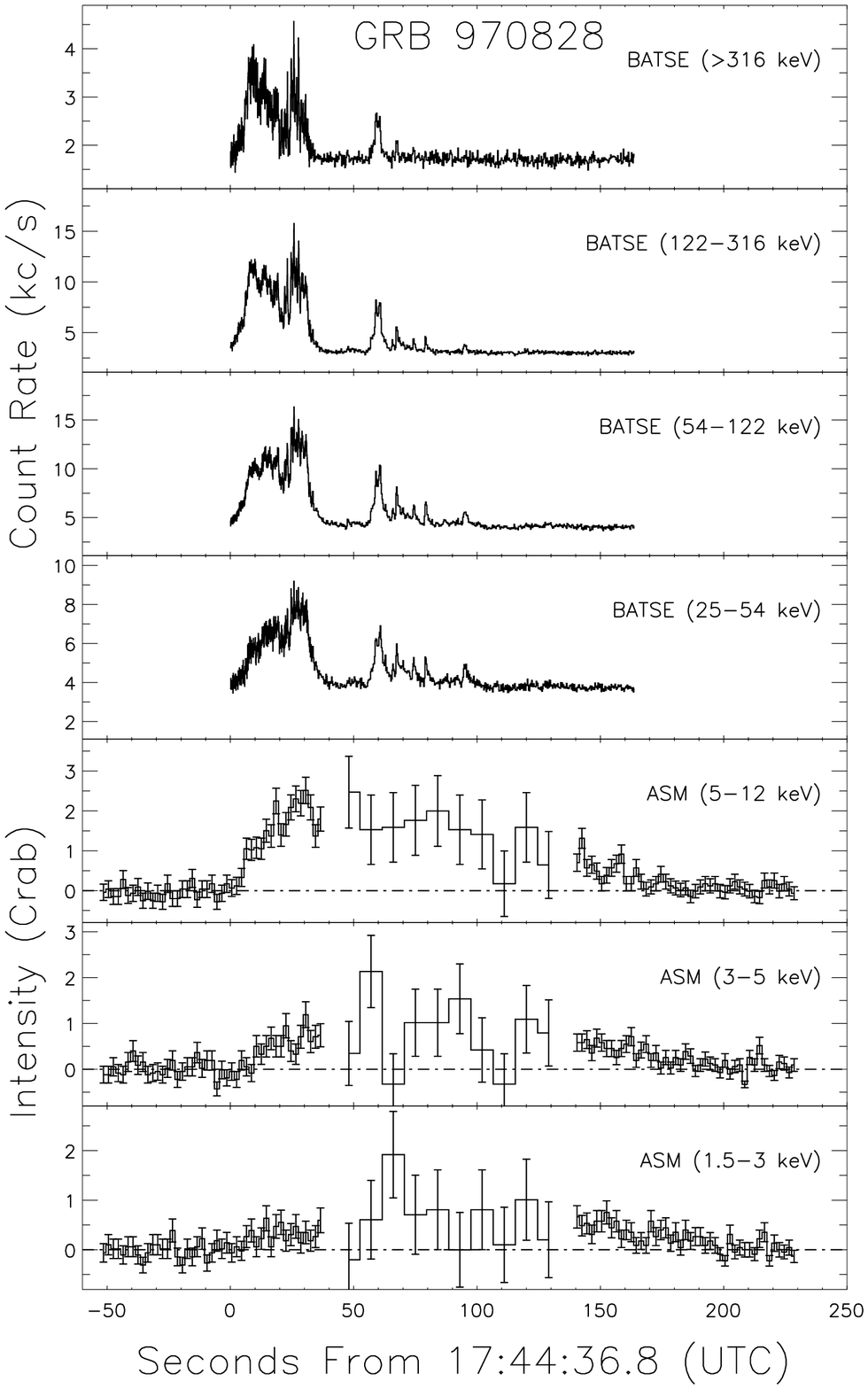 hoffset=-25 voffset=-14 hscale=58 vscale=58}{8.8cm}{14.1cm}
{\\\\\small Fig. 10 -- ASM time-series data in 2-s and 9-s bins for
GRB~970828, compared with data from BATSE.  The burst was first
observed with SSC~1, and with SSC~2 during the second and third
dwells.}
\label{fig:lc970828}

\pfg
\noindent
duration of the X-ray event is longer than its gamma-ray counterpart.

GRB~980703 (Fig.~\ref{fig:lc980703}) was detected in the FOVs of both
SSCs~1 and 2 simultaneously.  The flux rose over $\sim30$~s to reach a
maximum measured value at the end of a 90-s dwell.  The hard X-ray
maximum, as measured by BATSE, leads the the soft X-ray maximum by at
least 6~s.  The rising part of the burst is more variable at higher
energies.  During the second dwell, the GRB source is only
$0\arcdeg.6$ from the edge of the FOV of SSC~1 and is out of the FOV
of SSC~2.  Both BATSE and the ASM detect a lengthy tail.  Since the
transition from burst to tail in BATSE occurred while the ASM was in
motion, we have no information on the X-ray properties of this
transition.

GRB~981220 (Fig.~\ref{fig:lc981220}) was observed in SSC~2 near the end of a
90-s dwell.  This is the brightest GRB yet observed in the ASM data, reaching a
flux of over 5~Crab in the Sum band (1.5--12~keV).  Although this event was not
observed by BATSE, it was observed in the GRBM on board \sax.  The X-ray flux
rises earlier than and

\refstepcounter{figure} 
\PSbox{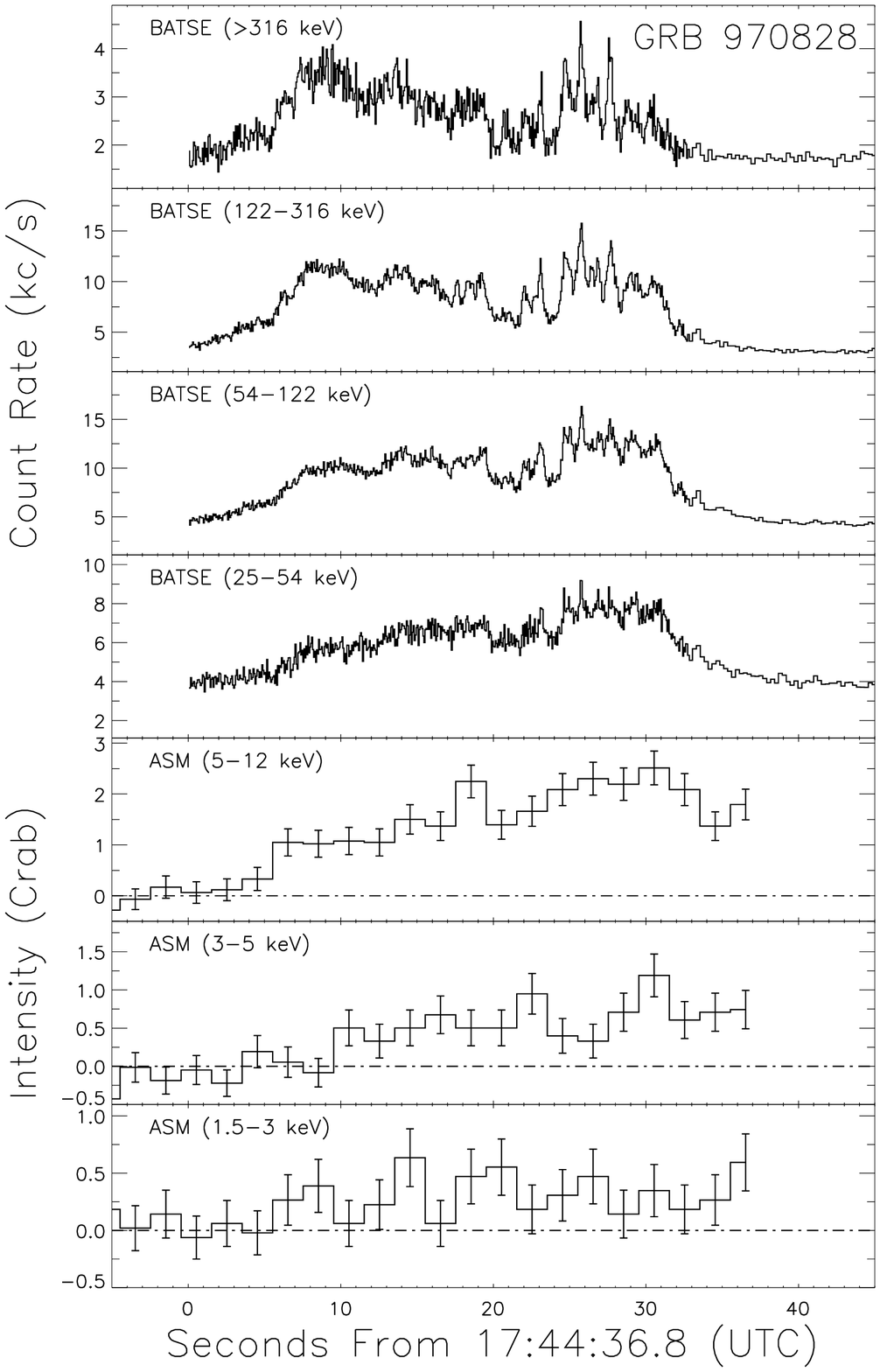 hoffset=-25 voffset=-14 hscale=58 vscale=58}{8.8cm}{14.1cm}
{\\\\\small Fig. 11 -- Light curves for GRB~970828
(Fig.~\protect{\ref{fig:lc970828}}) for the interval spanning 5~s before to
45~s after the BATSE trigger time.  There is a six-second gap at 37~s in the
lower panels because the ASM assembly was in motion.}
\label{fig:lc970828_c}

\pfg
\noindent
declines after the gamma-ray event, and the spectrum softens during the decay.

GRB~990308 (Fig.~\ref{fig:lc990308}) was observed in SSC~3 alone. An
optical transient is associated with this burst~\citep{schaf99}, so
the source position is accurately known.  BATSE also detected this
burst, and its count rate reveals a multiply-peaked light curve, some
35~s in duration.  The soft X-ray light curve does not appear to last
significantly longer than the hard X-ray light curve.

GRB~000301C (Fig.~\ref{fig:lc000301c}) was observed in SSC~2.  It consisted of
a single peak that reached over 3~Crab in the 5--12~keV band.  No emission was
detected in the 1.5--3~keV band, with a normalized $2\sigma$ upper limit of
90~mCrab (integrated over 90-s).  This burst was also detected by \uly~and {\it
NEAR}~\citep{shc00}.  The Earth lay between BATSE and the GRB at the time of
the event (M. Kippen, private communication), and the \sax~GRBM was powered
down due to the passage of the satellite near the South Atlantic Anomaly.

\refstepcounter{figure} 
\PSbox{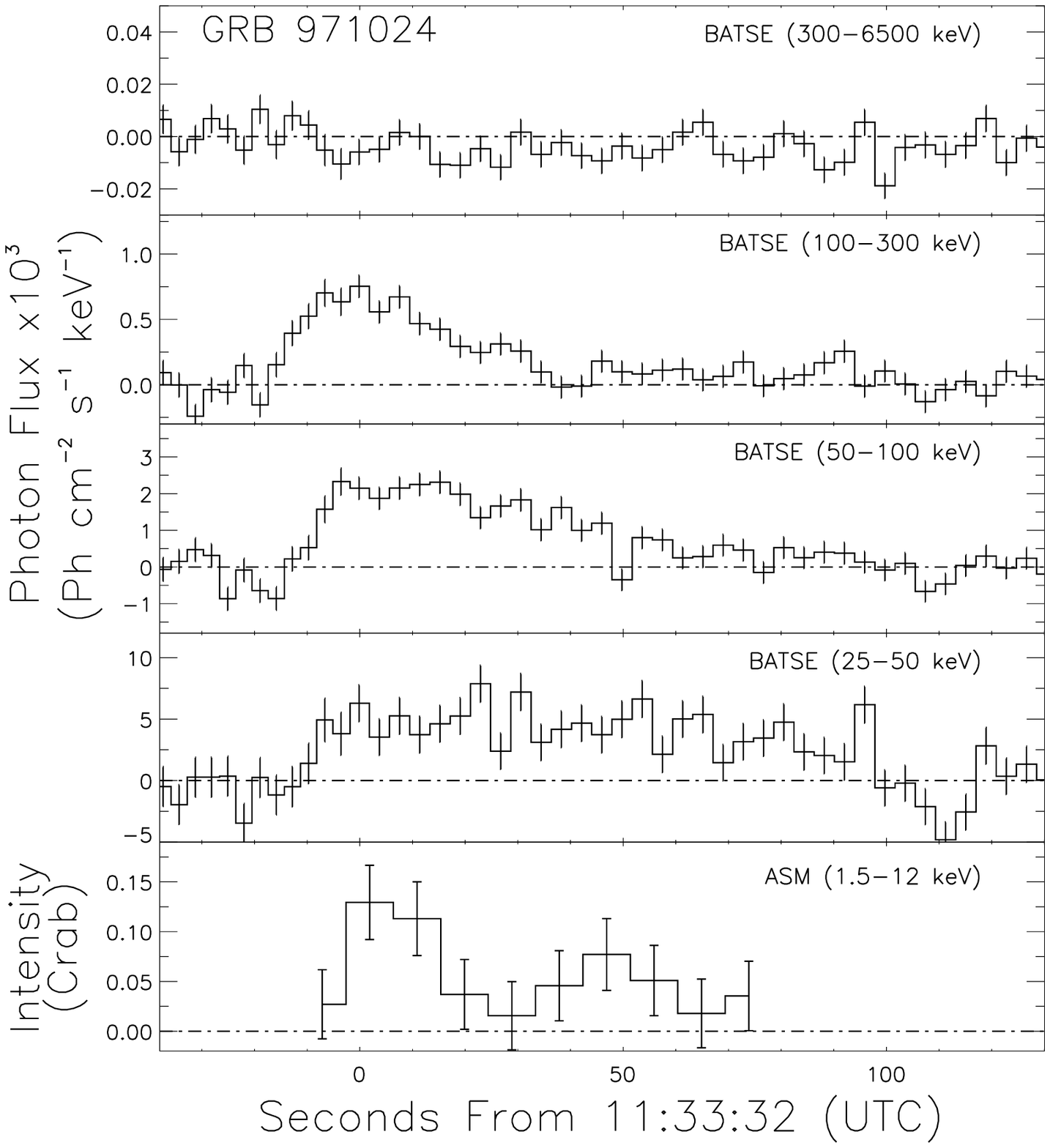 hoffset=-23 voffset=-11 hscale=55 vscale=55}{8.8cm}{9.9cm}
{\\\\\small Fig. 12 -- Time-series data for GRB~971024 in SSC~1 (9-s bins;
1.5--12~keV) and four BATSE energy bands (3.8-s bins).}
\label{fig:lc971024}

\refstepcounter{figure} 
\PSbox{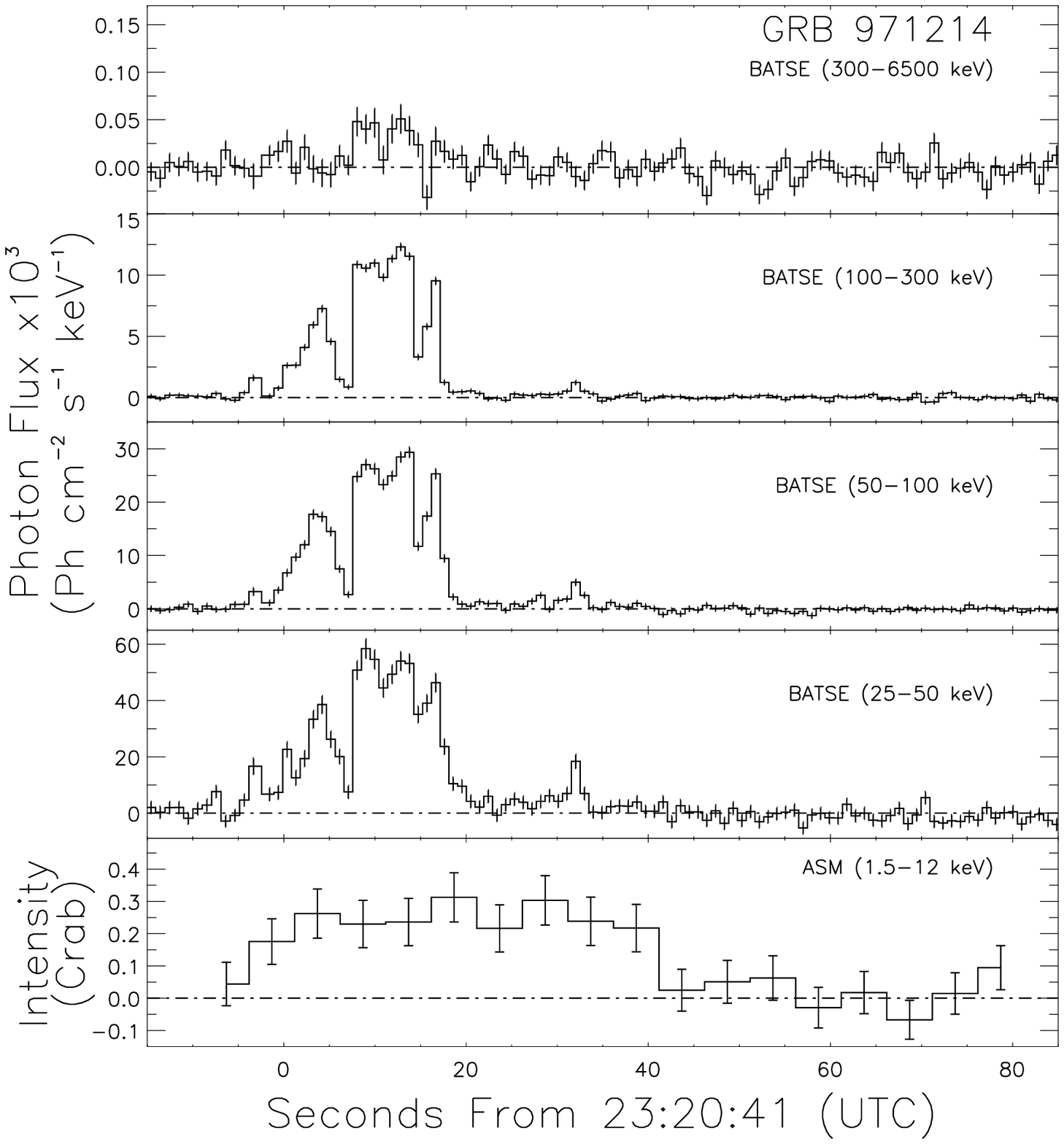 hoffset=-25 voffset=0 hscale=55 vscale=55}{8.8cm}{11.5cm}
{\\\\\small Fig. 13 -- Time-series data for GRB~971214 in SSC~3 (5-s bins;
1.5--12~keV) and four BATSE energy bands (0.64-s bins).}
\label{fig:lc971214}

\refstepcounter{figure} 
\PSbox{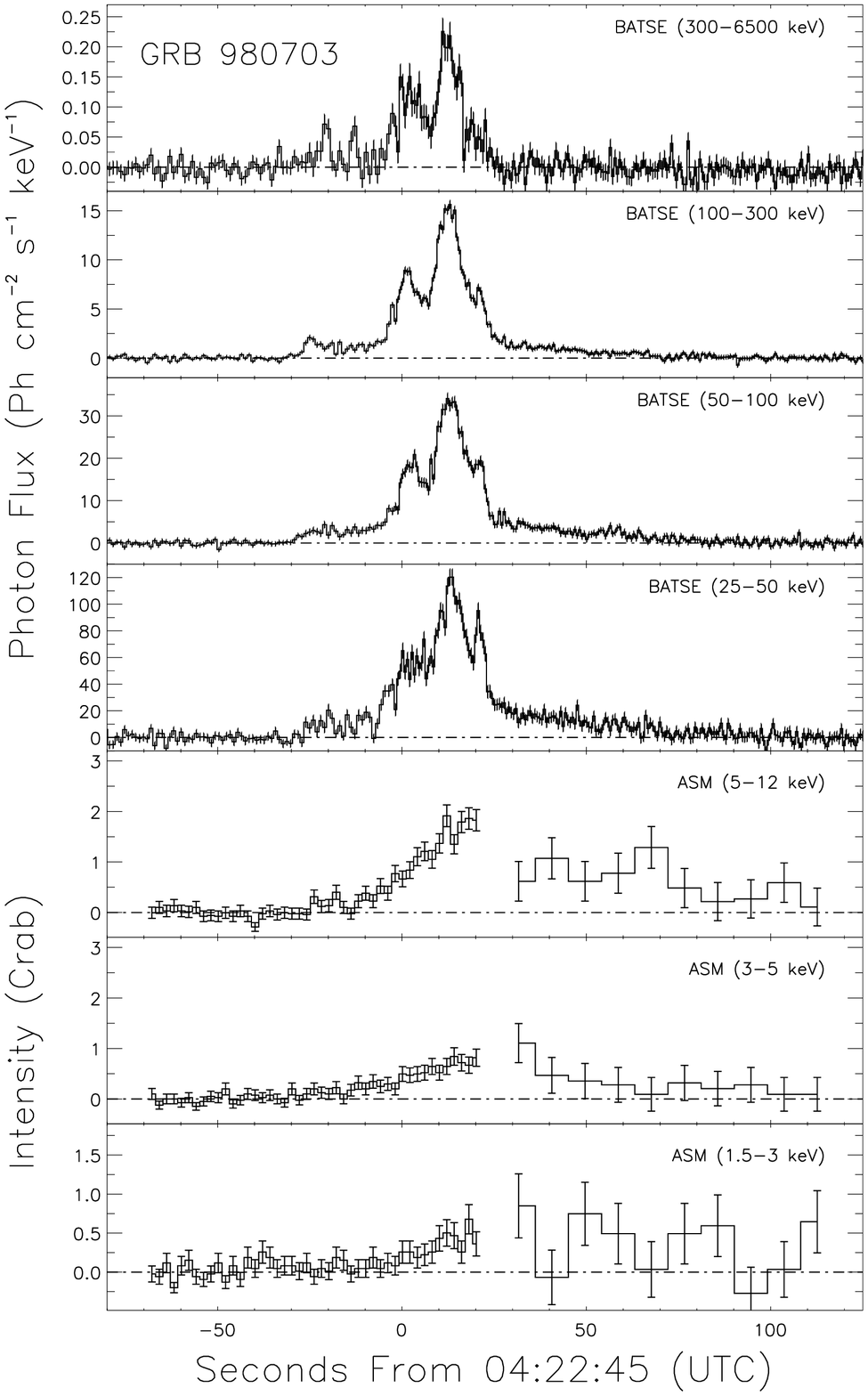 hoffset=-25 voffset=-10 hscale=55 vscale=55}{8.8cm}{14.3cm}
{\\\\\small Fig. 14 -- Time-series data for GRB~980703 both in SSC~1
and SSC~2 and later in SSC~1 alone (2-s and 9-s bins; 1.5--12~keV) and
four BATSE energy channels (0.64-s bins after the trigger, 1-s bins
before the trigger; 25--6500~keV).}
\label{fig:lc980703}

\pfg

\section{DISCUSSION\label{sec:lcsum}}

The external shock model has proven popular for explaining the major
features of most GRB afterglow behavior, and the internal shock model
has become the favored explanation for the temporal structure of the
GRB itself.  A clear observational distinction between the GRB event
and afterglow remains elusive, although several candidate criteria
have been proposed~\citep{ch98,gpkcw99,zhpf99}.  In this section, we
discuss the implications of these fifteen light curves for burst
origins in the context of the simple model outlined in
section~\ref{sec:grbmod}.  We discuss whether or not the observations
are consistent with an origin in synchrotron radiation, and in three
cases, we address the question of whether or not the ASM has observed
the onset of the X-ray afterglow.  A detailed analysis of spectral
evolution in bursts observed with the ASM is beyond the scope of this
paper.

The GRB light curves presented here are diverse in form.  Very few of
these bursts appear as simple fast-

\refstepcounter{figure} 
\PSbox{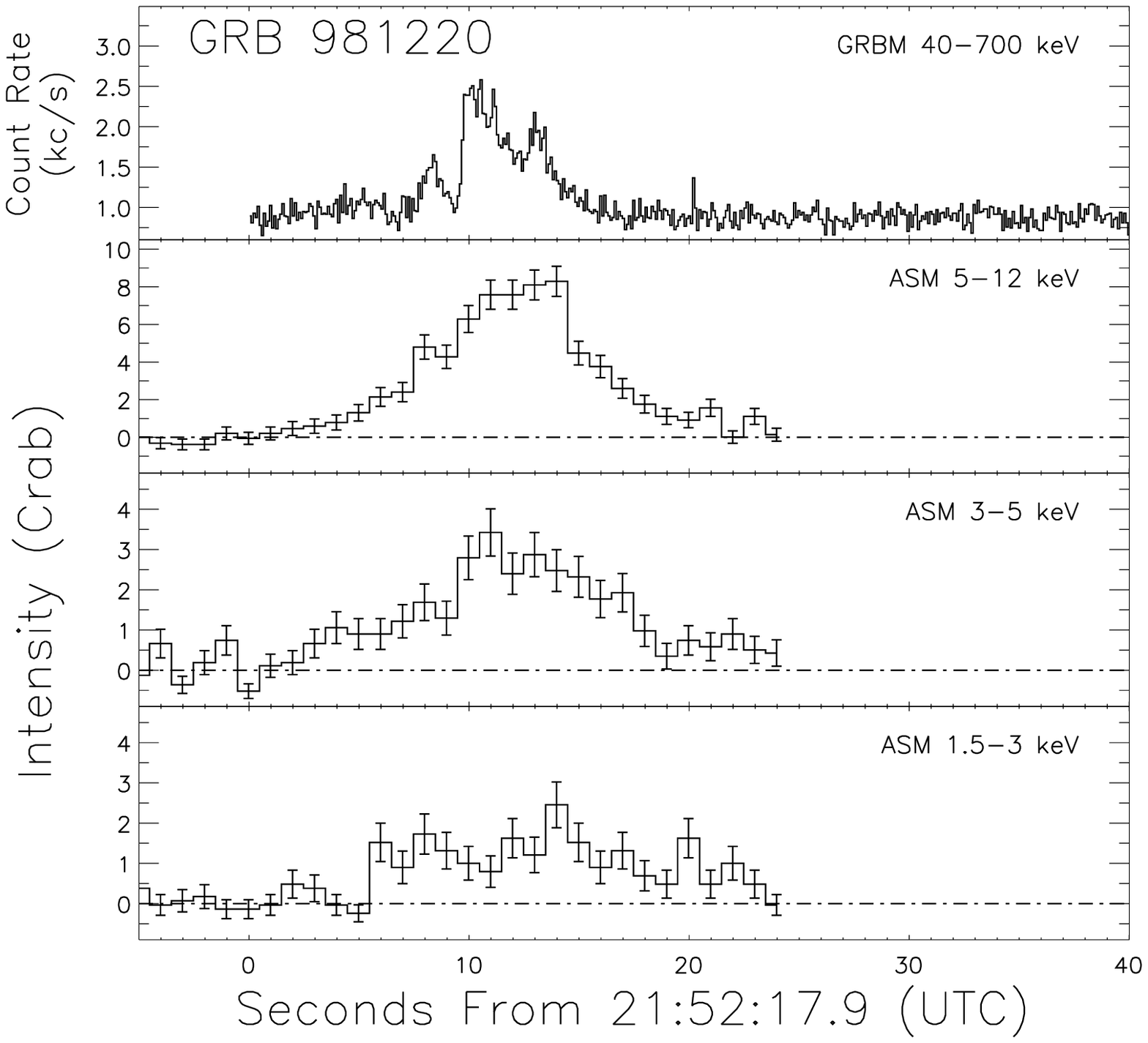 hoffset=-25 voffset=-10 hscale=55 vscale=55}{8.8cm}{8.2cm}
{\\\\\small Fig. 15 -- Time-series data (1.5--12~keV) for the ASM
SSC~2 observation of GRB~981220 in 2~s bins and the count rate in the
\sax~GRB Monitor (40--700~keV).  The spike at 20~s is an artifact.}
\label{fig:lc981220}

\pfg
\noindent
rise, exponential-decay shapes that one might expect from impulsive events.  In
X-rays, several bursts show a near-symmetric single-peak ({\it
e.g.}~Fig.~\ref{fig:lc981220}).  Some bursts show significant structure on
smaller time scales ({\it e.g.} Fig.~\ref{fig:lc960416}), while others do not
({\it e.g.} Fig.~\ref{fig:lc000301c}).  For many bursts, the substructure is an
energy-dependent phenomenon ({\it e.g.}  Figs.~\ref{fig:lc960529}
and~\ref{fig:lc980703}).  Two bursts (Fig.~\ref{fig:lc961002}
and~\ref{fig:lc961216}) even seem to have a slow-rise, fast-decay structure!
About half the bursts show multiple, distinct peaks, while five
(Figs.~\ref{fig:lc960727}, \ref{fig:lc961002}, \ref{fig:lc970828},
\ref{fig:lc971214}, and~\ref{fig:lc990308}) have a veritable forest of peaks
within their gamma-ray light curves.

Although each of these bursts shares some characteristics across the
entire observed energy range, there are other features that are only
detected in a few, or even one, energy channels.  The second gamma-ray
peak of GRB~960529 (Fig.~\ref{fig:lc970815}) is smeared out at low
energies and difficult to recognize as a distinct event.  This
smearing at low energies is common, and it can be seen in the onset of
GRB~980703 (Fig.~\ref{fig:lc980703}) as well as GRB~960727
(Fig.~\ref{fig:lc960727}), GRB~961002 (Fig.~\ref{fig:lc961002}), and
GRB~981220 (Fig.~\ref{fig:lc981220}).  In contrast, GRB~960416
(Fig.~\ref{fig:lc960416}) showed a peak unique to the softest ASM
energy channels.

GRB~970815 (Fig.~\ref{fig:lc970815}) shows multiple peaks, but in this
case the last peak is unusually soft.  An intriguing possibility is
that the third peak is due to an external shock, and hence represents
the beginning of the afterglow, while the first two peaks originate in
internal shocks that occur before the outermost ejecta sweep up enough
matter to instigate the external shock.  It has been predicted that
the afterglow could begin tens of seconds after the
burst~\citep{sari97,sp99}.  We address the possibility that the ASM
has detected the onset of X-ray afterglow emission by examining here
the three bursts for which searches for X-ray afterglow were carried
out hours after the event.

If the third peak in GRB~970815 does represent the on-

\refstepcounter{figure} 
\PSbox{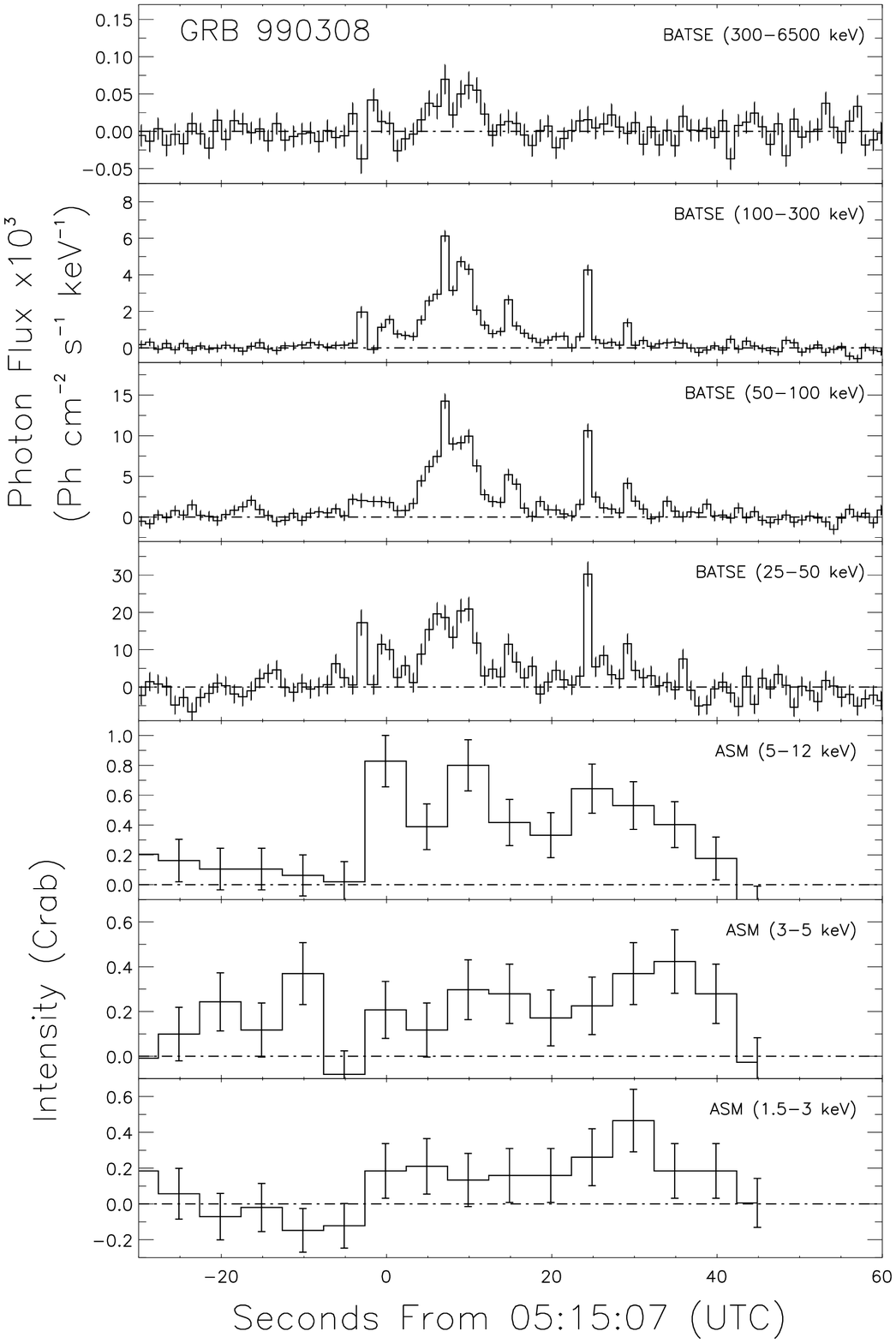 hoffset=-20 voffset=-6 hscale=52 vscale=52}{8.8cm}{13.7cm}
{\\\\\small Fig. 16 -- Time-series data for GRB~990308 in SSC~3 (5-s bins;
1.5--12~keV) and BATSE (1-s bins; 25--6500~keV).}
\label{fig:lc990308}

\pfg
\noindent
set of the afterglow, the soft lag would be the result of the decay of $\nu_m$
(see Eq.~\ref{eq:peaklimgam}), which is predicted by the external shock model
of~\citet{mr97} to fall as $(t-t_{\rm o})^{-2/3}$, where $t$ is measured by a
distant observer at rest with respect to the blast center and $t_0$ is the
initiation time, when $\nu_m$ is infinite.  There is a $\sim8$~s difference
between the times when the third peak reached its maximum in the C band
($E\sim7$~keV) and the A band ($E\sim2.25$~keV) at $t \sim 154$~s and 162~s,
respectively.  If this delay is due to the evolution of $\nu_m$, the shock that
generated the third peak must have begun cooling at 152~s after the BATSE
trigger time.  We therefore use this time as $t_0$ to model the subsequent
evolution of the emission.

The decay curve of the third peak can be fit with a power-law model, such that
$F(t) \propto (t-t_0)^{-\beta}$.  All three ASM energy bands show a decay from
the third peak consistent with $\beta = 1.3\pm0.1$.  This achromatic decay is
in marked contrast to the decay from GRB~970828, as described below, but it is
consistent with the afterglow observations from other GRBs and the predictions
of the external shock model~\citep{mr97}.  X-ray afterglow curves have been
measured for twenty-three GRBs that occurred prior to 1999 August, and the
power-law in-

\refstepcounter{figure} 
\PSbox{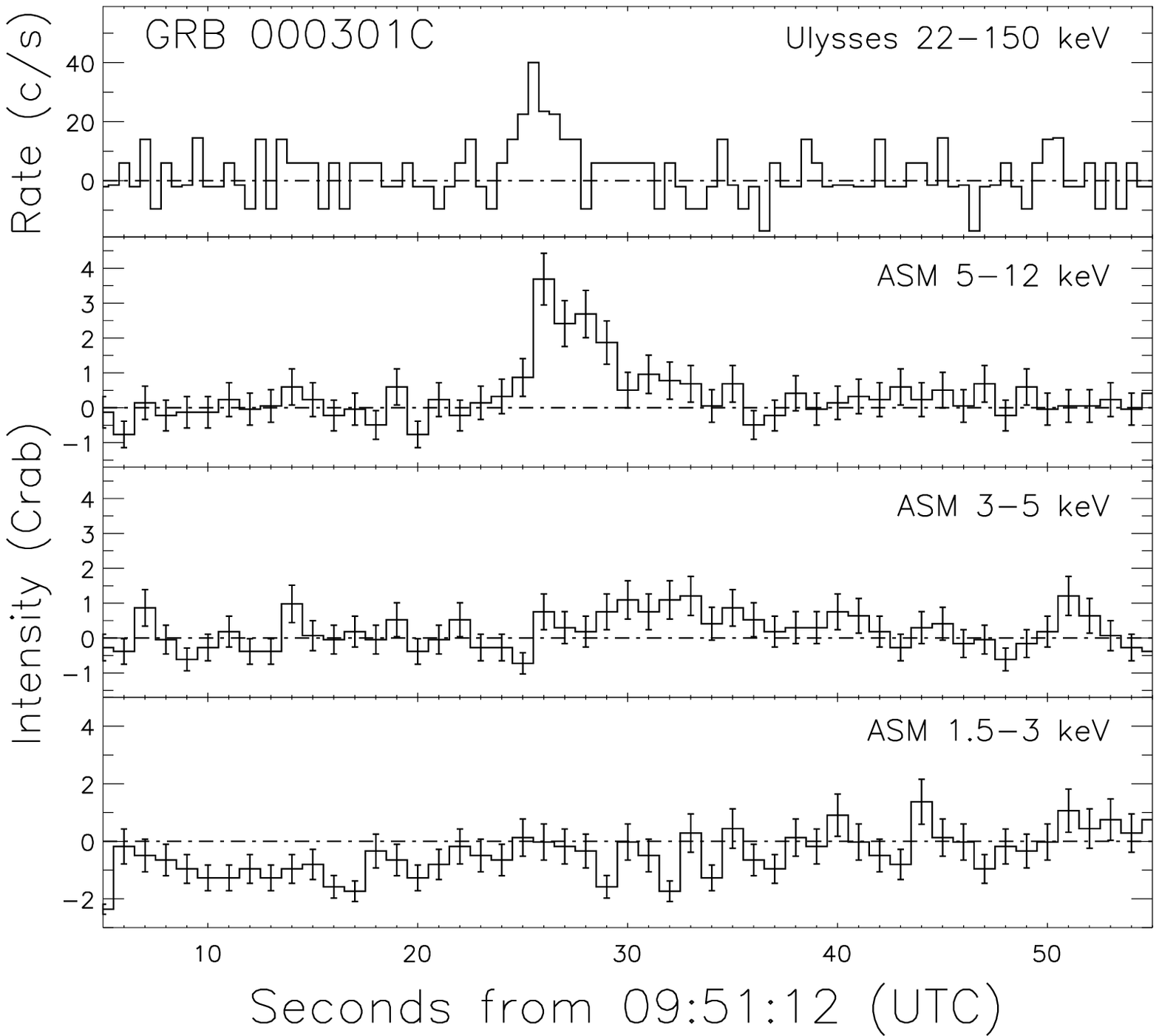 hoffset=-20 voffset=-6 hscale=52 vscale=52}{8.8cm}{7.75cm}
{\\\\\small Fig. 17 -- GRB 000301C as observed in SSC~2 (1-s bins;
1.5--12~keV) and the GRB detector on \uly~(0.5-s bins; 22--150~keV).
The rising count rate in the 1.5-3~keV channel indicates that the
\rxte~may be moving into a region of high activity in the Earth's
magnetosphere, or there may be interference from scattered Solar
X-rays.}
\label{fig:lc000301c}

\pfg
\noindent
dices for the decay range from 1.1 for GRB~970508~\citep{paabc98} to 1.57 for
GRB~970402~\citep{naabc98}.  The decay from GRB~970815 is thus fully consistent
with an afterglow-type decay.  If one extrapolates this decay to the time of
the \asca~follow-up observation, $\sim3.5\times10^5$~s after $t_0$, the
predicted 2--10~keV flux of about $8\times10^{-15}$~ergs~cm$^{-2}$~s$^{-1}$
lies below the \asca~upper limit of
$10^{-13}$~ergs~cm$^{-2}$~s$^{-1}$~\citep{muify97}.  The lack of an
\asca~detection thus does not rule out the possibility that the third peak is
the start of an afterglow decay; neither does it support that interpretation.

An alternative scenario is that the third peak results from a shell
catching up with the decelerating external shock; an early version of
the enhancements seen late in the afterglow of other bursts, such as
GRB~970508~\citep{paabc98}.  This would indicate that whatever
processes produce bursts continue to operate throughout the entire
event.  If this scenario is true, the afterglow and the burst cannot
always be considered distinct events.  It is possible, however, that
the late pulses in the afterglow are not due to collisions from
behind, as interpreted by~\citet{paabc98}, but instead are isolated
instances of the remnant colliding with a dense patch of external
medium.

GRB~970828 (Fig.~\ref{fig:lc970828}) displays an extended interval of
X-ray emission beyond the cessation of gamma-ray activity.  As in the
case of GRB~970815, we can ask if the ASM X-ray light curve reveals
the onset of the X-ray afterglow.  The final X-ray decay should then
share temporal and spectral properties with X-ray afterglow observed
at later times~\citep{facmp00}.  A power law decay curve, with the
origin set at the BATSE trigger time, when fit to the 1.5--12~keV band
data in the last ASM dwell, indicates that the flux decays as roughly
$t^{-5}$ (Fig.~\ref{fig:ag0828}).  All observed afterglow decay curves
have indices between $\sim1.1$~and $\sim1.6$, and the X-ray decay from
this GRB was

\refstepcounter{figure} 
\PSbox{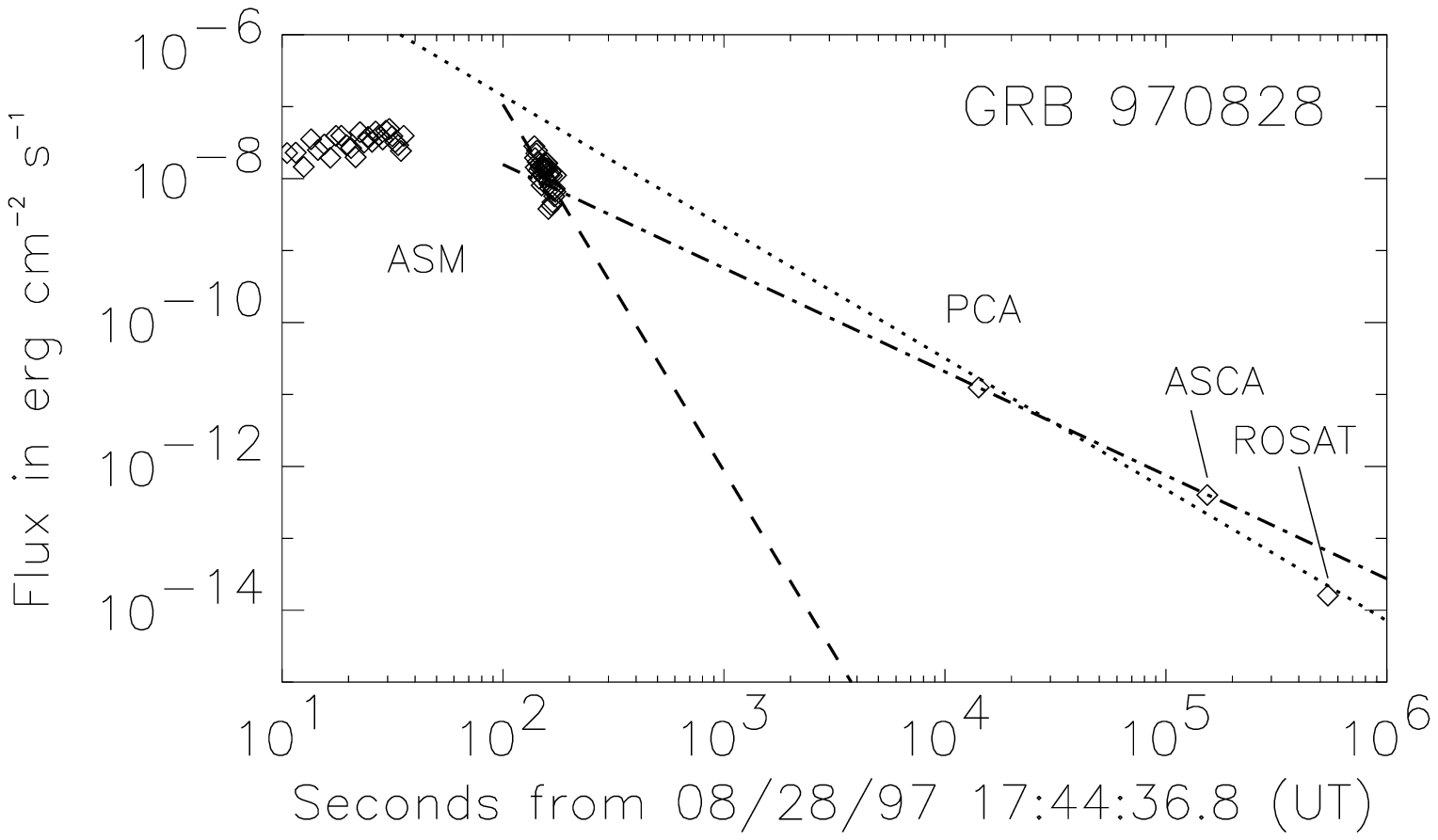 hoffset=-20 voffset=-6 hscale=58 vscale=58}{8.8cm}{5.7cm}
{\\\\\small Fig. 18 -- X-ray flux history of GRB~970828
($\sim$2--10~keV), as measured by the ASM, the
PCA~\protect{\citep{mcc97}}, \asca~\protect{\citep{muykm97}}, and
\rosat~\protect{\citep{gsgg97}}.  The dashed line shows the best-fit
power-law decay curve for the ASM data, $F\propto t^{-5}$.  The dotted
line shows the best-fit power-law for the three late-time flux
measurements, $F\propto t^{-0.5}$.  The ROSAT flux is derived by
extrapolating the reported 0.5--2.4~keV spectrum out to 10~keV.  If
this extrapolation is excluded from consideration, the PCA and
\asca~data are consistent with a power law decay of index
1.4~\protect{\citep{ynokm98}}, shown here as a broken line.}
\label{fig:ag0828}

\pfg 
\noindent
measured by the PCA and \asca~over the following two days to decay as
$t^{-1.4}$~\citep{muykm97}, as shown in Figure~\ref{fig:ag0828}.  There is
nothing in the theory of external shocks to explain a decay index around 5 that
later changes to 1.4.  Although the assignment of $t_0$ to the BATSE trigger
time is somewhat arbitrary, the best-fit decay index is inconsistent with, and
steeper than, a value of 1.4 for any value of $t_0$.  We therefore find it
unlikely that the ASM data represent the afterglow.

A fading X-ray afterglow was associated with GRB~980703
(Fig.~\ref{fig:lc980703}) through observations with the \sax~Narrow
Field Instruments (NFI) 22~h after the event~\citep{vgoog99}, and in
contrast to GRB~970828, the tail of the ASM decay curve and the NFI
flux measurements are consistent with a single power-law decay
curve~(Fig.~\ref{fig:ag0703}).  The best-fit decay index is $\sim1.3$,
a typical value for GRB afterglows.  This value is consistent with the
lower limit of 0.9 measured using only the NFI
observations~\citep{vgoog99}.  The tail of the burst emission may
therefore represent a transition to the afterglow.  However, the large
errors in the ASM measurements would allow the NFI flux measurements
to vary by orders of magnitude and still appear consistent with a
single decay.  It is worth noting that the BATSE count rates reveal a
second interval of emission from this burst, roughly 300~s after the
onset of the event, at which time the burst position was outside the
ASM FOV.  The apparent connection between the decay from the first
peak and the \sax~afterglow is therefore likely a coincidence.

Even beyond these three examples, the X-ray flux from GRB events tends to be
less variable and in almost all cases lasts longer than the associated
gamma-ray flux.  These increased peak widths may be examined in the context of
synchrotron radiation theory.  The cooling timescale for an electron undergoing
synchrotron energy loss is inversely

\refstepcounter{figure} 
\PSbox{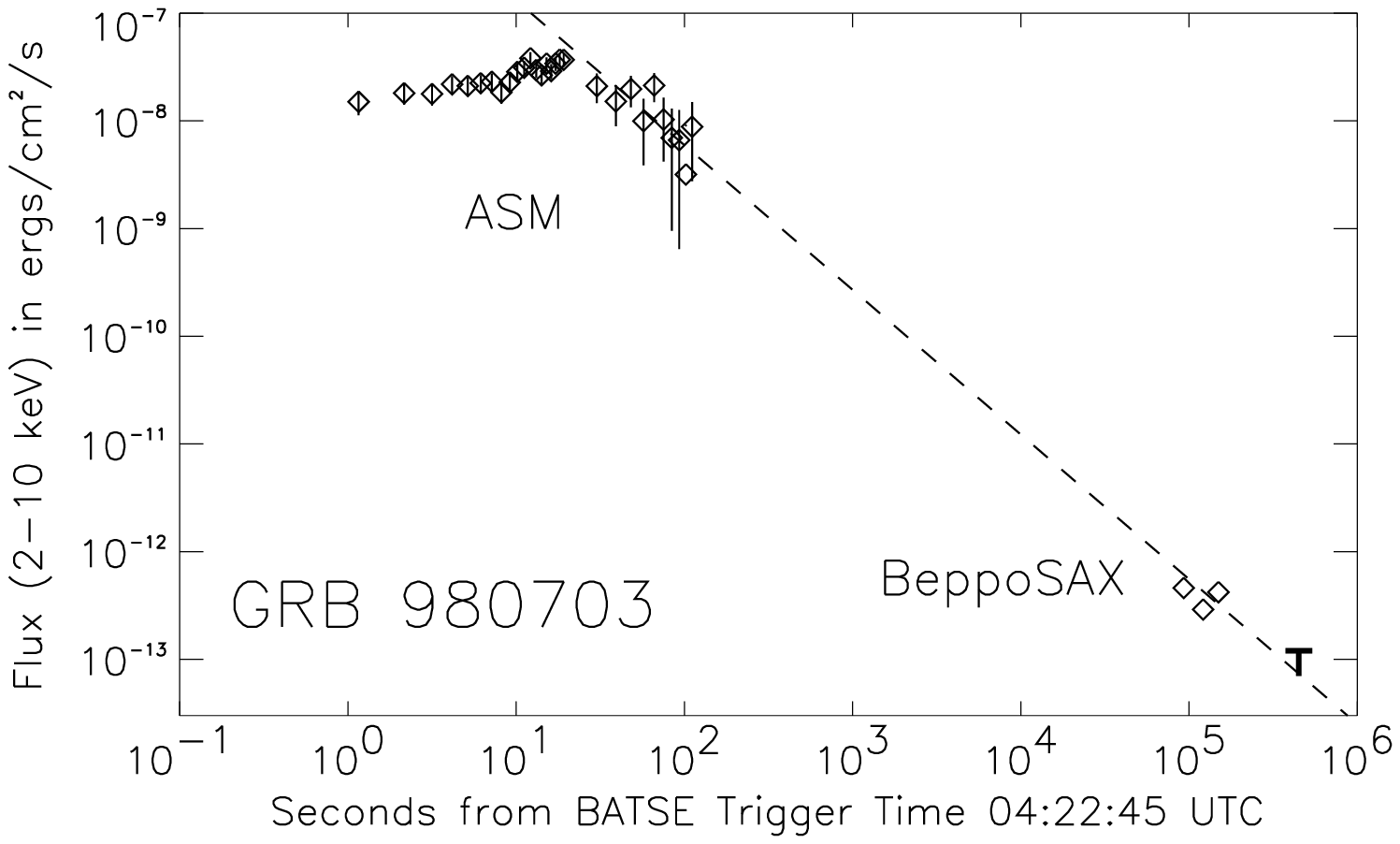 hoffset=-20 voffset=-6 hscale=58 vscale=58}{8.8cm}{5.6cm}
{\\\\\small Fig. 19 -- X-ray flux history of GRB~980703
($\sim$2--10~keV), as measured by the ASM and the
\sax~NFI~\protect{\citep{vgoog99}}.  The ``T'' shape indicates an
upper limit.  The dashed line shows the best-fit power-law decay curve
combining both instruments, $F\propto t^{-1.3}$.}
\label{fig:ag0703}

\pfg
\noindent
proportional to the electron's Lorentz factor ($t_c \propto 1/\gamma_e$).  The
synchrotron frequency of an emitting electron (and the characteristic photon
energy $E$ of the emitted radiation) goes as the square of its Lorentz factor
($E \propto \gamma_e^2$).  Hence, if a population of electrons is cooling
through synchrotron radiation, one would expect to find the cooling time
scales, or peak widths, to vary as $E^{-1/2}$~\citep{rl79,piran99,wg99}.
Although geometric effects can extend the time over which a burst is visible to
a distant observer, the relationship between the cooling time scale and the
width of observed peaks in a GRB light curve is expected to be
preserved~\citep{piran99}.

The fact that the bursts presented here were observed serendipitously
by different instruments renders the width of the GRB peaks difficult
to measure in many cases.  The time resolution of the ASM MTS data,
exacerbated by the relatively small effective area, also makes
comparison of individual burst features problematic ({\it e.g.}
Figs.~\ref{fig:lc960727} and~\ref{fig:lc961002}).  Nevertheless, we
have measured peak widths as a function of energy band for seven
bursts, and we display the results in Figure~\ref{fig:widvse}.  Here,
a peak width is estimated as the best-fit Gaussian function, with an
uncertainty given by the $2\sigma$ confidence interval for the
Gaussian width.  We also measured an exponential decay timescale,
yielding equivalent results.  For bursts with complex temporal
structure, like GRB~970828 (Fig.~\ref{fig:lc970828}) above 25~keV, a
Gaussian function is an extremely poor match to the shape of the light
curve, but we cite the best-fit Gaussian in order to compare duration
consistently with the simpler bursts.

Figure~\ref{fig:widvse} shows that only the simplest bursts are
consistent with the prediction of synchrotron cooling.  For example,
the widths for GRB~000301C (Fig.~\ref{fig:lc000301c}), with a
single-peaked Gaussian shape in both the ASM and \uly~data, are
consistent with the expected power-law solution with index of $-0.5$
(Fig.~\ref{fig:widvse}h, dashed line).  While GRB~960416
(Fig.~\ref{fig:lc960416}) displayed multiple peaks, these peaks were
widely separated, and the width of each peak is consistent with a
$E^{-1/2}$ scaling law (Fig.~\ref{fig:widvse}a \&~b).  Again, the
dashed line in the figure shows a representative power law with index
$-0.5$.

\refstepcounter{figure} 
\PSbox{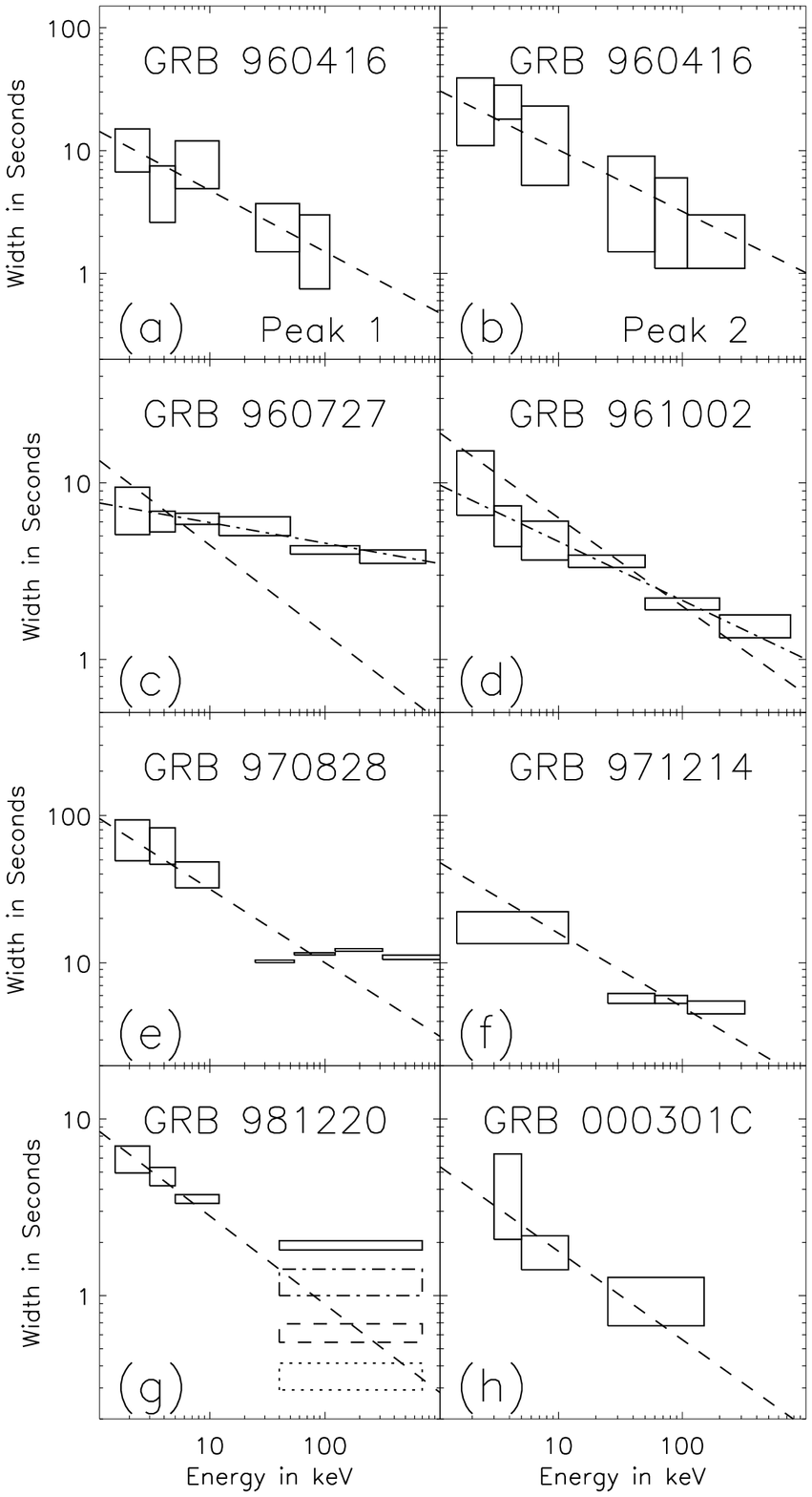 hoffset=-17 voffset=-54 hscale=72 vscale=72}{8.8cm}{16.5cm}
{\\\\\small Fig. 20 -- Peak width vs. energy for seven GRBs.  The GRB modeled
with a simple Gaussian function, and the $\pm2\sigma$ confidence interval for
the width is plotted on the $y$-axis, against the appropriate energy channel.
Data are from BATSE, the \sax~GRBM, \uly, Konus, and the ASM.  The dashed lines
in each plot represent the $E^{-0.5}$ dependence expected from synchrotron
cooling.  The broken lines in panels (c) and (d) indicate the best-fit slopes.}
\label{fig:widvse}

\pfg

Such a simple model, however, does not fit any of the other bursts.
GRB~960727 (Fig.~\ref{fig:lc960727}) and GRB~961002
(Fig.~\ref{fig:lc961002}) seem simple as recorded by the ASM, but the
Konus light curves reveal complex temporal structure.  The total
duration of both these bursts has a much flatter dependence on energy
than synchrotron cooling would predict; the best-fit power law indices
are inconsistent with a slope of $-0.5$ (Compare the broken and dashed
lines in Figures~\ref{fig:widvse}c and d).  GRB~981220
(Fig~\ref{fig:lc981220}) also displayed a simple, single-peaked
structure in the ASM time-series data, but the \sax~count rates
resolve this into three narrower peaks at high energies.  The ASM data
alone are consistent with a power law index of $-0.5\pm0.1$
(Fig.~\ref{fig:widvse}g), but the extrapolation of this scaling law to
the energy range of the \sax~GRBM is inconsistent with the measured
width of the peak as modeled by a single Gaussian (the solid box
spanning the 40--700~keV range).  However, if the GRBM light curve is
modeled by the superposition of three Gaussian events, then their
widths are (in chronological order) $0.35\pm0.1$~s, $0.6\pm0.1$~s, and
$1.2\pm0.2$~s (shown as the dotted, dashed, and broken boxes,
respectively).  These three boxes are all consistent with the
extrapolated power-law, and perhaps the emission seen in soft X-rays
is dominated by the cooling of a single shock that produced one of
these high-energy peaks.

The widths of the ASM light curves for GRB~970828
(Figs.~\ref{fig:lc970828} \&~\ref{fig:widvse}e) and GRB~971214
(Figs.~\ref{fig:lc971214} \&~\ref{fig:widvse}f) do not match well with
an extrapolation of the primary peak width as observed by BATSE.  In
both cases, the BATSE light curves showed multiple peaks of emission,
and a Gaussian model of the primary peak widens much more slowly than
a $E^{-0.5}$ power law would predict.  Within the context of the
internal shock model, the extended length of the ASM light curve for
GRBs~970828 and 971214 can perhaps be interpreted as the reheating of
the matter heated by the initial shock, which does not have time to
cool in the X-ray regime before it is shocked again.  However, there
is no reason why the multiple shocks would necessarily be shocking the
same electrons.  This scenario also would not explain the smaller
indices for GRBs~960727 and 961002 (Fig.~\ref{fig:widvse}c \& d).

In short, only the simplest bursts display the $E^{-0.5}$ dependence
of width on energy predicted by synchrotron cooling.  We point out
that the only other burst for which this hypothesis has been tested
across X-ray and gamma-ray bands, GRB~960720, was also a simple burst
with a single peak~\citep{phjcf98}.  A likely explanation of this
discrepancy is that multiple peaks are indicative of complex
interactions that violate the assumption of a single infusion of
energy followed by cooling through radiation.  It is unclear why some
complex bursts lead to abnormally long soft X-ray light curves (such
as GRB~970828), while others (like GRB~960727) show a much weaker
dependence on energy.  It is possible that individual peaks in these
complex bursts do behave consistently with the predictions of
synchrotron cooling, as do the well-separated peaks in GRB~960416 and
perhaps the short peaks in GRB~981220, but most often, the statistics
and time resolution of the ASM data do not allow us to track the
behavior of individual short peaks.

When study of individual peaks is possible, however, several show a
soft lag in their times of maximum count rate.  It is worth noting
that several do not, at least within the limitations of the effective
time resolution.  In no case did we observe any candidate for a
distinct X-ray precursor, such as that associated with
GRB~980519~\citep{zhpf99} or perhaps GRB~900126~\citep{minpf91}.
Precursor events are rare, so their absence in the ASM sample is
unsurprising.

What is perhaps surprising is the absence of GRBs shorter than 10~s in
duration.  The GRBs in the BATSE catalog have a well-known bimodal
duration distribution with peaks at 0.1~s and
10~s~\citep{hurl92,kmfbb93,fmwbh94,kpkpp96}.  Our variability search
was conducted on time-scales of 1/8~s, 1~s, and 9~s, and yet all the
GRBs we found are drawn from the longer sub-population, although
GRB~000301C is a borderline case~\citep{jfghh00}.  If the typical peak
intensities of the short bursts are of the same magnitude as or less
than those of the long bursts, the ASM is less likely to detect the
short bursts, although a short burst should still stand out in the
time-series data.  Our search was able to detect the short bursts from
SGR~1627--41~\citep{sbl99}.  The population of short bursts represents
roughly 25\% of the first BATSE GRB catalog~\citep{kmfbb93}, so
perhaps the absence of short GRB events in the ASM sample is a
statistical fluctuation, but it is noteworthy that all of the bursts
localized to date by the \sax~WFC have also been from the population
of longer bursts~\citep{gsccd00,facmp00}.  The BATSE data suggest that
the shorter bursts have harder spectra, so perhaps they are not bright
enough in the 1.5--12~keV range for ASM detection.

\acknowledgements

This project combined results from several instruments, and hence
could not have been completed without the help of many individuals.
Of crucial help was Scott Barthelmy's work in creating and maintaining
the GCN.  We would also like to acknowledge the support of the
\rxte~team at MIT and NASA/GSFC.  Support for this work was provided
in part by NASA Contract NAS5--30612.  D. A. Smith is supported by NSF
fellowship 00-136. KH is grateful for Ulysses and IPN support under
JPL Contract 958056 and NASA grants NAG 5--9503 and NAG 5--3500.

\clearpage

\newcommand{\noopsort}[1]{} \newcommand{\printfirst}[2]{#1}
  \newcommand{\singleletter}[1]{#1} \newcommand{\switchargs}[2]{#2#1}

\end{document}